\documentclass[rmp,aps,showpacs,twocolumn,nofootinbib,amsmath,amssymb,longbibliography,floatfix]{revtex4-1}

\usepackage{graphicx}
\usepackage{amsmath}
\usepackage{amsthm,amsmath,amsfonts,dsfont,bbm}
\usepackage{hyperref}
\usepackage{color}

\newcommand{\comment}[1]{}

\newcommand{\tr}[2][]{\text{Tr}_{#1}\left\{#2\right\}}
\newcommand{\trtxt}[2][]{\text{Tr}_{#1}\{#2\}}
\newcommand{\ketbra}[1]{\mathopen{|}#1\mathclose{\rangle}\hspace{-0.25em}\mathopen{\langle}#1\mathclose{|}}
\newcommand*{\bra}[1]{\mathopen{\langle}#1\mathclose{|}}
\newcommand*{\ket}[1]{\mathopen{|}#1\mathclose{\rangle}}
\newcommand*{\braket}[2]{\mathopen{\langle}#1{|}#2\mathclose{\rangle}}
\newcommand{\beq}{\begin{equation}}
\newcommand{\eeq}{\end{equation}}
\newcommand{\beqa}{\begin{eqnarray}}
\newcommand{\eeqa}{\end{eqnarray}}

\definecolor{augustine}{rgb}{0,0,0}
\newcommand{\ak}{\textcolor{augustine}}
\definecolor{hendrik}{rgb}{0,0,0}
\newcommand{\hw}{\textcolor{hendrik}}
\definecolor{roman}{rgb}{0,0,0}
\newcommand{\ro}{\textcolor{roman}}

\begin{document}
	
	\title{Simulation methods for open quantum many-body systems}
	
	\author{Hendrik Weimer}
	\email{hweimer@itp.uni-hannover.de}
	\affiliation{Institut f\"ur Theoretische Physik, Leibniz Universit\"at Hannover, Appelstra{\ss}e 2, 30167 Hannover, Germany}
	
	\author{Augustine Kshetrimayum}
	 \affiliation{Dahlem Center for Complex Quantum Systems, Physics Department, Freie Universit\"{a}t Berlin, 14195 Berlin, Germany}

	\author{Rom\'an Or\'us}
	\affiliation{Donostia International Physics Center, Paseo Manuel de Lardizabal 4, E-20018 San Sebasti\'an, Spain}
	\affiliation{Ikerbasque Foundation for Science, Maria Diaz de Haro 3, E-48013 Bilbao, Spain}

	\begin{abstract}
		
		
		Coupling a quantum many-body system to an external environment
		dramatically changes its dynamics and offers novel possibilities not
		found in closed systems. Of special interest are the properties of
		the steady state of such open quantum many-body systems, as well as
		the relaxation dynamics towards the steady state. However, new
		computational tools are required to simulate open quantum many-body
		systems, as methods developed for closed systems cannot be readily
		applied. We review several approaches to simulate open many-body
		systems and point out the advances made in recent years towards the
		simulation of large system sizes.

	\end{abstract}
	
	
	\maketitle
	
	\tableofcontents
	
	
	\section{Introduction}
	
	\subsection{The open quantum many-body problem}
	
	Open quantum many-body systems have witnessed a surge of
        interest in recent years, chiefly for two reasons. On the one
        hand, these systems offer the exciting possibility to use
        controlled dissipation channels to engineer interesting
        quantum many-body states as the stationary state of their
        dynamics \cite{Diehl2008,Verstraete2009,Weimer2010}. On the
        other hand, open quantum many-body systems are attractive from
        a fundamental perspective, as their dynamics exhibits a wide
        range of features not found in equilibrium systems. As in the
        case of closed quantum systems, the complexity of the problem
        scales exponentially with the size of the system, requiring
        the use of sophisticated simulation methods to obtain useful
        results.
	
	Interestingly, open quantum many-body systems are even harder
        to simulate on classical computers than closed systems, while
        at the same time the stationary state of an open quantum
        system is much easier to experimentally prepare than the
        ground state of a closed system. These properties make open
        quantum systems one of the prime candidates to show a quantum
        advantage of quantum simulators over classical methods within
        noisy intermediate-scale quantum devices
        \cite{Preskill2018}. However, this requires a thorough
        assessment of the capabilities of classical simulation
        methods, which we will provide in this review.

        In our review, we first provide a general introduction to open
        quantum many-body systems, laying particular emphasis on the
        key differences compared to simulating closed quantum systems
        and on the paradigmatic models that have emerged to benchmark
        simulation methods for open systems. In the main part, we
        first review stochastic methods commonly known as
        wave-function Monte-Carlo techniques, which are based on a
        numerical exact treatment of the total Hilbert space of the
        problem. We then turn to tensor network simulation techniques
        aiming to describe the ``physical corner'' of the Hilbert
        space, i.e., the quantum states that are most relevant to
        describe the dynamical evolution and steady states of open
        quantum many-body systems. Subsequently, we review variational
        methods that employ very similar strategies, including
        variational methods that are based on a tensor network
        description.  We \ak{also} cover phase space methods and closely
        related counterparts. \ro{Finally, we have added a section on linked cluster expansion.} Within our review, we will not cover
        methods derived from a field-theoretical description of open
        quantum systems within the Keldysh formalism, as this has
        already been extensively covered in a previous review article
        \cite{Sieberer2016}.  \ak{We will also not cover integrable models \cite{Prosen_quadraticfermion,Prosen_betheansatz,Maghrebi_solvablefamily,Prosen_exactMPO,Prosen_exactMPO2,Guoexact}
        	for which analytical techniques such as the Bethe ansatz can be employed.}

        \subsection{The Markovian quantum master equation}

        The state of an open system is described by its density
        operator $\rho$, which can be described as a statistical ensembles of pure states,
        \begin{equation}
          \rho = \sum\limits_i p_i\ketbra{\psi_i},
        \end{equation}
        where $p_i$ denotes the probability to find the system in the
        state $\ket{\psi_i}$. Note that the decomposition into pure
        states is not unique. In our review, we will limit ourselves
        to the discussion of Markovian systems, i.e., dynamical
        systems in which the generator of the dynamics
        $\mathcal{L}[\rho]$ (commonly called the Liouvillian) depends
        only on the state at the present time $t$ and not on the state
        at earlier times. Such Markovian systems form a dynamical
        semigroup and can be described by a quantum master equation in
        Lindblad form
	\begin{equation}
		\begin{split}
			\frac{d}{dt}\rho &= \mathcal{L} [ \rho] \\ 
			&= -i\left[H,\rho \right] + \sum_{\mu}\left( L_\mu \rho L_\mu^\dagger - \frac{1}{2}L_\mu^\dagger L_\mu \rho - \frac{1}{2} \rho L_\mu^\dagger L_\mu \right) , 
			\label{master0}
		\end{split}
	\end{equation}
	where $H$ is the Hamiltonian of the system and $\{ L_\mu,
        L_\mu^\dagger \}$ the Lindblad operators responsible for the
        incoherent dynamics arising from the coupling to an external
        environment, which are also known as the jump operators
        \cite{Gorini1976,Lindblad1976}.

        The validity of the Lindblad master equation
        Eq.~(\ref{master0}) for a concrete physical system
        crucially depends on the separation of several
        timescales. Considering a system of interest coupled to a
        larger environment, one first assumes a weak coupling between
        system and environment, such that the entanglement between
        system and environment remains low. Furthermore, the
        environment must not retain any memory of the system degrees
        of freedom. The approximations related to these conditions are
        commonly refered to as the Born-Markov approximation
        \cite{Breuer2002} and require that the correlation time of the
        environment $\tau_E$ is much smaller than the relaxation time
        of the system $\tau_R$. Finally, the differences in
        eigenfrequencies in the system $\omega_s$ has to be large
        compared to the inverse relaxation time $\tau_R^{-1}$.

        These approximations are well justified in quantum optical
        systems, in particular atoms coupled to electronically excited
        states
        \cite{Raitzsch2009,Baumann2010,Barreiro2011,Krauter2011,Malossi2014}. There,
        the optical frequencies of the transition leave a large time
        scale to observe the complete relaxation to its equilibrium
        state. Additionally, the relaxation of the electronic
        excitation into the vacuum of the radiation field as the
        correlation time of the radiation field is related to the
        photon frequency \cite{Breuer2002}, which again is much larger
        than the relaxation rate $\tau_R^{-1}$. Artificial atomic
        systems such as the nitrogen-vacancy center in diamond
        \cite{Jelezko2004,Dutt2007,Robledo2011} offer similar benefits.

        Another advantage of quantum optical systems for studying open
        quantum many-body systems is the possibility to drive them
        with time-dependent laser fields. Importantly, if all the jump
        operators in the master equation describe transitions between
        the eigenstates of the Hamiltonian, the resulting steady state
        of the system is guaranteed to be a thermal state
        \cite{Breuer2002}. However, if an oscillatory driving term is
        added to the system Hamiltonian, it is possible to observe
        non-equilibrium steady states in the rotating frame of the
        driving. Optically excited atoms can also exhibit strong
        interactions when excited to Rydberg states
        \cite{Saffman2010}, which can be used to realize a rich
        variety of driven-dissipative quantum many-body systems
        \cite{Lee2011,Glatzle2012,Ates2012a,Lemeshko2013a,Rao2013,Carr2013a}.
        
         \ro{There are also interesting solid state platforms to study strong interaction and dissipation. One example is that of semiconductor polaritonic systems see, e.g., \cite{Carusotto2013}), where semiconductor microstructures are used to embed quantum wells or quantum dots, becoming a photonic resonator where strong interactions can be induced. Another example is that of circuit-QED systems \cite{Fitzpatrick2017,Ma2019}, where superconducting circuits can be used to construct Bose-Hubbard lattices of microwave photons, and where dissipation can be engineered, so that one can have a tailored reservoir.}

         \ro{It is important to remark that the Lindblad operators are usually considered to be local, but this approximation holds only in the weak-coupling limit. To be more precise, a Markovian master equation with quasi-local Lindblad operators holds as long as the coupling between the system and the environment is weak, which in practice amounts to (1) a slow development of correlations between system and environment, (2) fast decay of excitations of the environment, and (3) neglect of fast-oscillating terms when compared to the typical system timescale. One should be careful, however, since when dealing with \emph{strongly correlated} systems,} \hw{strong interactions within the system of interest may lead to a breakdown of the local Lindblad dissipation \cite{Wichterich2007,Beaudoin2011}. In these cases, it may be necessary to consider additional steps to derive the correct Lindblad operators \cite{Reiter2012}. For the purpose of our review, we assume that the correct Lindblad form has already been derived.}

	\subsection{Steady state solution versus time evolution}
	
	Typically, there are two different aspects that are of interest when
	studying open quantum many-body systems. First, one wishes to
	understand the properties of one or several steady states that the
	system reaches in the long time limit. This is similar to
	understanding the ground state properties of a closed many-body
	system. Second, one is interested in the dynamical evolution of the
	system towards the steady state. The latter is particularly
	interesting when the system exhibits several steady states that can be
	reached depending on the initial condition of the system.

        \hw{While the requirements for the appearance of a unique
          steady state are well understood for finite systems
          \cite{Spohn1976}, many-body systems add the additional
          complication that the long time limit and the thermodynamic
          limit do not necessarily commute.} In some cases, even when
        chiefly interested in the steady state, it is more efficient
        to compute the full time evolution of the system. This is
        comparable to imaginary time evolution algorithms to find the
        ground state of a closed many-body system. In our review, we
        will contrast the two approaches and address this distinction
        when discussing individual simulation methods in the main part
        of our review.

        Investigating the full time evolution also offers the
        possibility to investigate interesting many-body effects
        during the relaxation dynamics. For instance, it is possible
        for open many-body systems exhibiting a quite trivial steady
        state, while the relaxation behavior is dominated by complex
        glassy quantum dynamics \cite{Olmos2012}.
	
	\subsection{Differences to equilibrium problems}

        To find the steady state of an open quantum many-body system,
        it might first be tempting to take well established methods
        for ground state calculations for closed systems and try to
        adapt it to the open case. Unfortunately, this approach fails
        in many cases. For example, quantum Monte-Carlo methods that
        are highly successful for ground state calculations, require
        to rewrite the partition function of the quantum system to a
        corresponding classical system. However, for the steady state
        of an open system it is unclear a priori (and often incorrect
        \cite{Sieberer2013}) whether the steady state of the system is
        a thermal state that can be described in terms of a partition
        function. The same argument holds for density functional
        theory approaches trying to minimize the ground state energy;
        usually, the steady state of an open system is completely
        different from the ground state of the Hamiltonian. This is
        even true in the limit of infinitely weak dissipation, as the
        strength of the dissipation will predominantly control the
        relaxation rate rather than the properties of the steady
        state.

        Some methods from the study of closed quantum systems
        out of equilibrium can be adopted to open systems; we will
        discuss these cases in detail. In general, the simulation of
        an open quantum system is computationally much harder than for
        a closed system due to the statistical nature of the state.

        \hw{Additionally, one can benefit to some extent from the vast
          body of works commited to the study of classical
          non-equilibrium dynamics. For example, the importance of the
          symmetries of the open quantum many-body dynamics is equally
          important as in the classical case \cite{Hohenberg1977} and
          allows for the classification of dissipative phase
          transitions in terms of their universality classes.}
	
	\subsection{Paradigmatic models}

        Within the analysis of ground state many-body problems, there
        is a number of particular models that have found especially
        wide interest and are often used as a first example to
        benchmark a numerical method. These models include, e.g., the
        Ising model in a transverse field, the Heisenberg model, and
        the Hubbard model (both bosonic and fermionic). A similar
        observation can be made about open quantum many-body problems,
        where these paradigmatic models are often derived from the
        corresponding ground state counterparts, i.e., the Hamiltonian
        dynamics is the same. However, adding dissipation to a closed
        many-body model can be done in different ways and can lead to
        drastically different results. In the following, we present
        and briefly discuss the two most promiment dissipative
        many-body models; we will provide a more detailed discussion
        in later sections when referring to particular numerical
        strategies to tackle them.

        One of the most widely studied open many-body models in recent
        years is the transverse field Ising model with longitudinal
        dissipation \cite{Lee2011}. Its Hamiltonian is of the form of
        the conventional Ising model, given in terms of Pauli matrices
        $\sigma_\alpha$ by
        \begin{equation}
          H =  \frac{h}{2} \sum\limits_i \sigma_x^{(i)} + \frac{V}{4} \sum\limits_{\langle ij\rangle}  \sigma_z^{(i)}\sigma_z^{(j)},
          \label{eq:dising}
        \end{equation}
        where $h$ is the strength of the transverse field and $V$
        accounts for the Ising interaction. The dissipation is
        incorporated in terms of jump operators of the form $c_i =
        \sqrt{\gamma}\sigma_-$, with $\gamma$ being the rate of
        dissipative flips from the spin up to the spin down state. An
        important aspect is that the dissipation breaks the $Z_2$
        Ising symmetry of the Hamiltonian, i.e., the quantum master
        equation does not exhibit such a symmetry. The model is also
        relevant to ongoing experiments in the field of interacting
        Rydberg atoms \cite{Carr2013,Malossi2014}.

        Within a mean-field calculation \cite{Lee2011}, the model is
        predicted to support a large range of $h$ values for which the
        system exhibits two stable steady states. We will discuss in
        later sections of our review how different numerical
        approaches address the question on the existence of such a
        bistable thermodynamic phase. \hw{According to mean-field theory,
        the bistable region ends in a critical point that belongs to
        the Ising universality class \cite{Marcuzzi2014}.}

	Another important dissipative model is the driven-dissipative
        Bose-Hubbard model. While there are different ways to
        generalize the famous Bose-Hubbard model \cite{Fisher1989} to
        the dissipative case, the most commonly studied one involves a
        dissipative particle loss that can be countered by a coherent
        driving term \cite{LeBoite2013, Carusotto2013}. Its
        Hamiltonian is given by
        \begin{equation}
          H = -J\sum\limits_{\langle i,j\rangle} b_i^\dagger b_j + \sum\limits_i \left[\frac{U}{2}n_i^2 - \Delta\omega n_i + F \left(b_i+b_i^\dagger\right)\right].
          \label{eq:dhubbard}
        \end{equation}
        In this model, $J$ describes the hopping of bosons between
        sites, while the on-site interaction $U$ involves the square
        of the density operator $n_i = b^\dagger_ib_i$. Furthermore,
        $\Delta\omega$ is the chemical potential for the bosons, and $F$
        describes the aforementioned coherent driving. Finally, the
        quantum jump operators capturing the loss of a single boson
        are given by $c_i = \sqrt{\gamma} b_i$. While the dissipation
        term also breaks the $U(1)$ symmetry of the conventional
        Bose-Hubbard model, here, the symmetry is already broken on
        the level of the Hamiltonian by the inclusion of the driving
        term $F$.

\begin{figure}[t]
          \includegraphics[width=\linewidth]{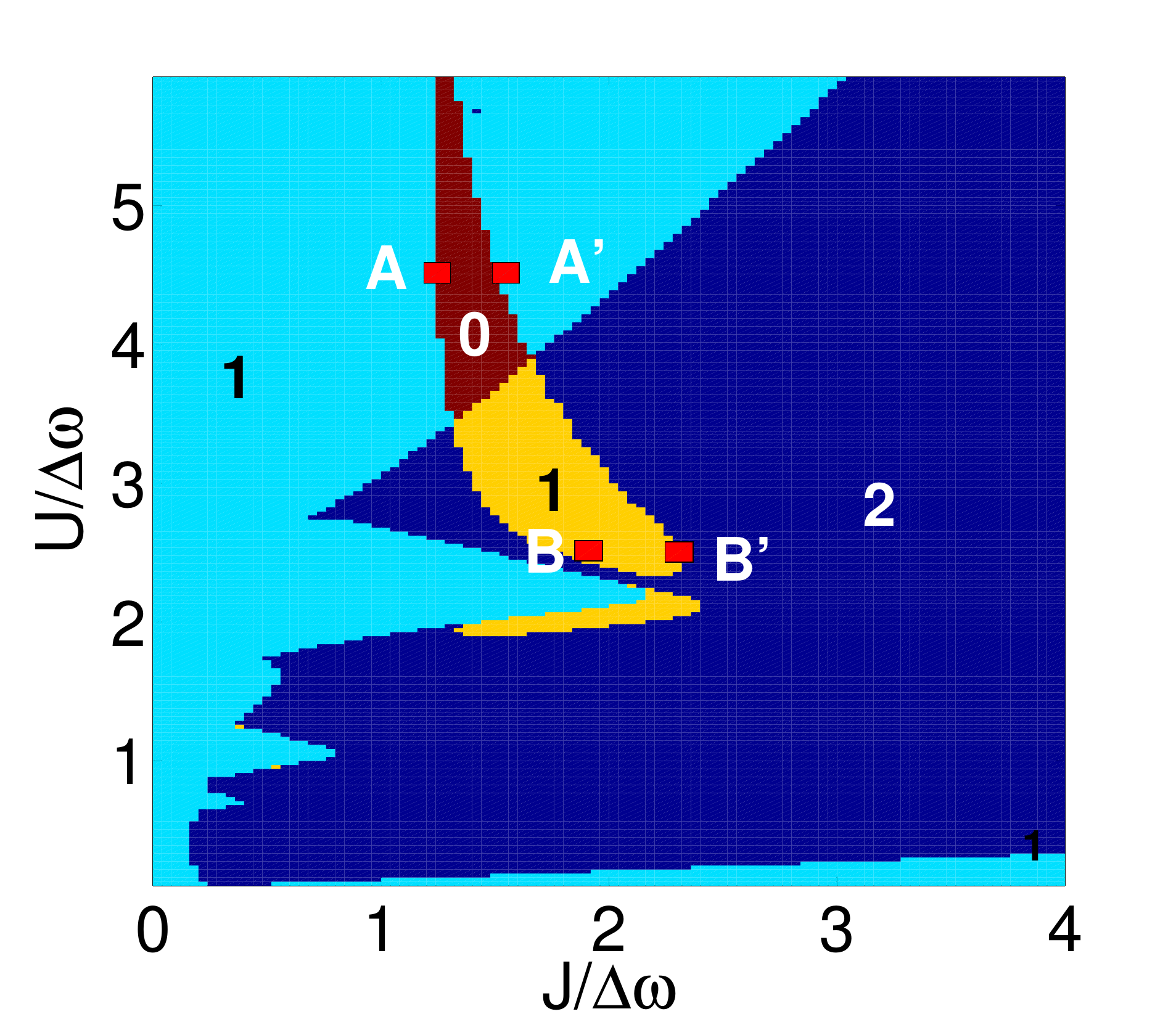}
          \caption{Mean-field phase diagram of the driven-dissipative Bose-Hubbard model. The numbers inside the plot represent the number of stable mean-field solutions. The yellow region exhibits two mean-field solutions, one of which is unstable. From \cite{LeBoite2013}.}
          \label{fig:mfbh}
        \end{figure}

        As with the dissipative Ising model, the driven-dissipative
        Bose-Hubbard model has a very intriguing mean-field phase
        diagram, where several islands of multistability occur in a
        way that is somewhat reminiscent of Mott lobes
        \cite{LeBoite2013}, see Fig.~\ref{fig:mfbh}. The stability of
        the mean-field solutions has been evaluated by considering density matrices of the form
        \begin{equation}
          \rho = \prod\limits_i (\rho_i^{MF}+\delta\rho_i),
        \end{equation}
        with $\rho_i^{MF}$ being the mean-field solution for the
        steady state. Expanding the quantum master equation up to
        first order in $\delta\rho_i$ allows to evaluate the stability
        by checking whether none of eigenvalues of the Liouvillian has
        a positive real part.
          
	\section{Stochastic methods}

        \label{sec:wfmc}

        Upon first glance, the computational complexity of an open
        quantum system in terms of the Hilbert space dimension $d$
        appears to be at least $O(d^2)$, as there are $O(d^2)$
        independent entries in the density matrix $\rho$. However, the
        density matrix at an initial time $t_0$ can be written as a
        statistical ensemble of pure states, $\rho(t_0) = \sum_i p_i
        \ketbra{\psi_i(t_0)}$. Instead of propagating the entire
        density matrix, the key strategy is to propagate the
        individual pure states $\ket{\psi_i}$ to the time $t$
        and then calculate observables according to
        \begin{equation}
          \langle O\rangle = \tr{O \rho} = \sum\limits_i
          p_i\bra{\psi_i} O \ket{\psi_i}.
        \end{equation}
        The probability distribution $p_i$ can then be sampled using
        standard Monte-Carlo techniques, which is why the approach is
        often called the wave function Monte-Carlo method. In
        practice, the most common strategy is to start from an initial
        pure state $\ket{\psi_0}$ and perform $M = 1/p_i$ numerical
        simulations. Since the trajectories $\ket{\psi_i}$ are
        independent from each other, the statistical error associated
        with the observable will behave as $\Delta O \sim
        1/\sqrt{M}$. The entire computational cost will be $O(Md)$,
        which is considerably lower than $d^2$ already for quite
        modest system sizes. Importantly, the requirement to repeat
        the simulation $M$ times results in the simulation time being
        significantly longer than for a comparable closed quantum
        system. Depending on the observable, $M\approx 1000$ is a
        reasonable choice to get the statistical error down to a few
        percent. For spin $1/2$ systems, this essentially means that
        the system sizes that can be studied in an open system consist
        of $\log_2 M \approx 10$ particles less than in a closed
        system.

        The central question is now how can a single trajectory
        $\ket{\psi_i}$ be propagated such that the ensemble of all
        trajectories satisfies $\rho(t) = \sum_i p_i
        \ketbra{\psi_i(t)}$. One possiblity is to describe the
        evolution of the density operator in terms of a quantum state
        diffusion approach \cite{Gisin1992,Percival1998}, in which the
        incoherent dynamics from the Lindblad operators is captured in
        terms of a stochastic Schr\"odinger equation,
        \begin{equation}
          d\psi_i(t) = -i H_{eff}\ket{\psi_i(t)}dt + \sum_j M_j\ket{\psi_i(t)}dW_j,
        \end{equation}
        where the $dW_j$ refer to Wiener increments. The effective
        Hamiltonian $H_{eff}$ describes the drift of the state vector
        in the Hilbert space,
        \begin{equation}
          H_{eff} = H + \sum_j 2\langle c_j^\dagger \rangle c_j - c_j^\dagger c_j - \langle c_j^\dagger\rangle \langle c_j\rangle.
        \end{equation}
        The diffusion operators $M_j$ describe the random fluctuations
        arising from each associated jump operator $c_j$,
        \begin{equation}
          M_j = c_j - \langle c_j\rangle.
        \end{equation}
        This stochastic Schr\"odinger equation conserves the norm of
        the state vector and can be solved by standard techniques for
        stochastic differential equations.
        
        An alternative strategy to propagate a single trajectory is
        the quantum jump method
        \cite{Dalibard1992,Dum1992,Molmer1993,Plenio1998}. This
        approach has been recently reviewed extensively in
        \cite{Daley2014}, so we will only cover the basic
        strategy. Within the quantum jump method, the dynamics is
        split into to parts. First, the state $\ket{\psi_i}$ is
        propagated under an effective non-Hermitian Hamiltonian
        $H_{NH}$,
        \begin{equation}
          H_{NH} = H - \frac{i}{2}\sum_j c_j^\dagger c_j.
        \end{equation}
        Once the norm of the state drops below a previously drawn
        random number $r$, a quantum jump occurs. Which quantum jump
        occurs is drawn from the probability distribution
        \begin{equation}
        p_j =
        \mathcal{N} \langle \psi_i|c_j^\dagger c_j|\psi_i\rangle,
        \end{equation}
        with $\mathcal{N}$ being a normalization factor. While the
        high order integration of $H_{NH}$ is straightforward, a high
        order simulation of the quantum jumps requires a more subtle
        identification of the time the jump operator needs to be
        applied. For instance, the popular QuTiP library
        \cite{Johansson2012,Johansson2013} uses a logarithmic secant
        method to numerically solve the equation $\langle \psi_i(t) |
        \psi_i(t)\rangle = r$ for the time $t$.
        
        No matter which approach is used to propagate a single
        trajectory, the computations can be highly parallelized since
        the trajectories are independent from each other by
        construction. Doing so, it is possible to simulate open
        many-body spin $1/2$ models with up to 20 spins
        \cite{Raghunandan2018}. The relatively small system sizes when
        compared to equilibrium problems demand the development of new
        data analysis techniques, e.g., concerning finite size scaling
        methods. One possiblity is to use anisotropic system sizes to
        obtain more data points for a reliable finite size scaling
        extrapolation. Close to a phase transition, the susceptibility
        $\chi$ of a system may be expressed as
        \begin{equation}
          \chi = N^\alpha \tilde{\chi}(\lambda),
          \label{eq:scaling}
        \end{equation}
        where $N$ is the number of particles and $\alpha$ is an
        exponent associated with the underlying phase transition
        \cite{Binder1989}. The reduced susceptibility $\tilde{\chi}$
        is only a function of the anisotropy $\lambda$ of the system
        and can be determined by symmetry considerations as well as
        numerical data \cite{Raghunandan2018}.

        \hw{The wave-function Monte-Carlo method has been used to analyze
        the one-dimensional dissipative Ising model of
        Eq.~(\ref{eq:dising}) \cite{Ates2012a,Hu2013}. While these
        works have not found a bistable phase as predicted by
        mean-field theory, a significant increase in the spin
        correlations has been reported in the same region
        \cite{Hu2013}. Additionally, finite size scaling of a similar
        two-dimensional model believed to lie in the same universality
        class as the dissipative Ising model has found evidence for a
        first order transition \cite{Raghunandan2018}.}

	\section{Tensor network methods}

        \label{sec:tensor}
	
	\subsection{One spatial dimension}

        \label{sec:tensor1d}
        
	We would, first, like to describe the important numerical techniques that has been developed for studying open quantum many-body systems using Matrix Product States (MPS) which is the one-dimensional ansatz of the tensor network (TN) family. MPS, by far, is the most successful and widely used ansatz in comparison to other ansatz of the Tensor network family, thanks to the success of the density matrix renormalization group (DMRG) \cite{dmrg1,dmrg2} and related techniques \cite{tebd1,tebd2}. Not only are its properties very well understood, contraction of MPS tensors can be done efficiently and exactly unlike the case for its higher dimensional counterparts \cite{Schuchpepshard,Eisertpepshard}. For these reasons, MPS have been used extensively producing extremely accurate results, however, mostly in the context of ground state calculations of many-body systems \cite{Schollwock2005}. For a detailed review on MPS and other tensor networks in general, we ask the readers to refer to \cite{OrusTNreview,EisertTNreview,Schollwock2011,RomanNaturereview,VerstraeteTNreview,CiracTNreview,BiamonteTNreview}. The application to open quantum systems, meanwhile, is more rare and there are only a few known approaches one can take for such systems.  Not only are open systems more computationally challenging (since we need to deal with matrices in place of vectors for the pure states), there are also several intrinsic bottlenecks such as the positivity, hermiticity in the numerical optimization of the density matrix. Nevertheless, many of the ideas in the pure state formalism have been successfully applied in the context of open systems \ak{using the concept of Matrix Product Operators (MPOs) or Matrix Product Density Operators (MPDOs) \cite{MPOreviewVerstraete,MPDOreviewVerstraete}. These appraches have also been used to study thermodynamic properties of 1d systems}. We discuss them below.
	
\begin{figure}
			\begin{center}
				\includegraphics[scale = 0.24]{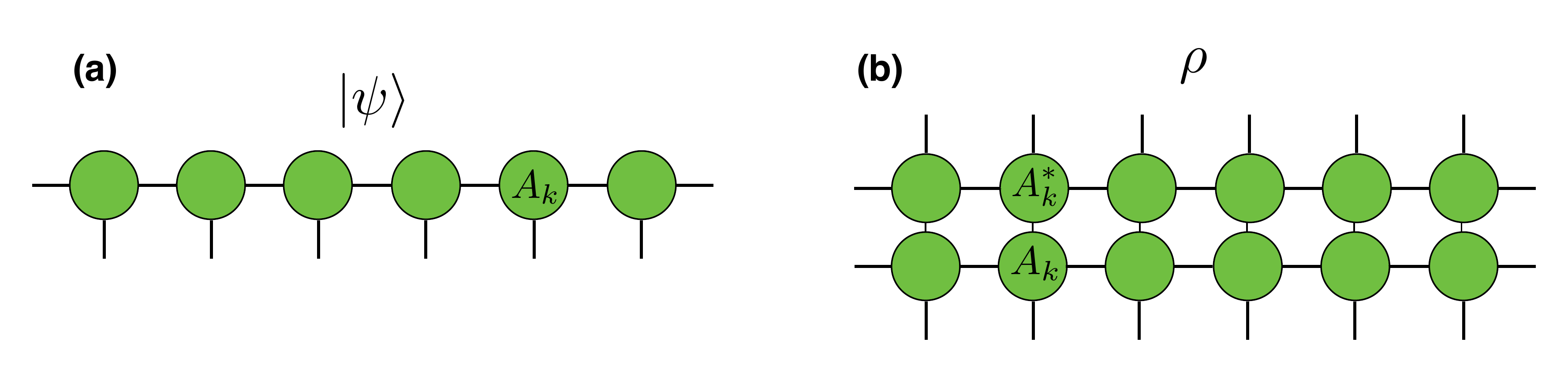}
				\caption{(a) Writing a wave function $|\psi \rangle$ as an MPS for 6 sites. Each site has a physical dimension $d$. (b) A density matrix $\rho$ can be written as an MPDO, an extension of the MPS formalism. Such a construction automatically ensures positivity of the density matrix.}
				\label{MPDO}
			\end{center}
		\end{figure}

\begin{figure}
			\begin{center}
				\includegraphics[scale = 0.24]{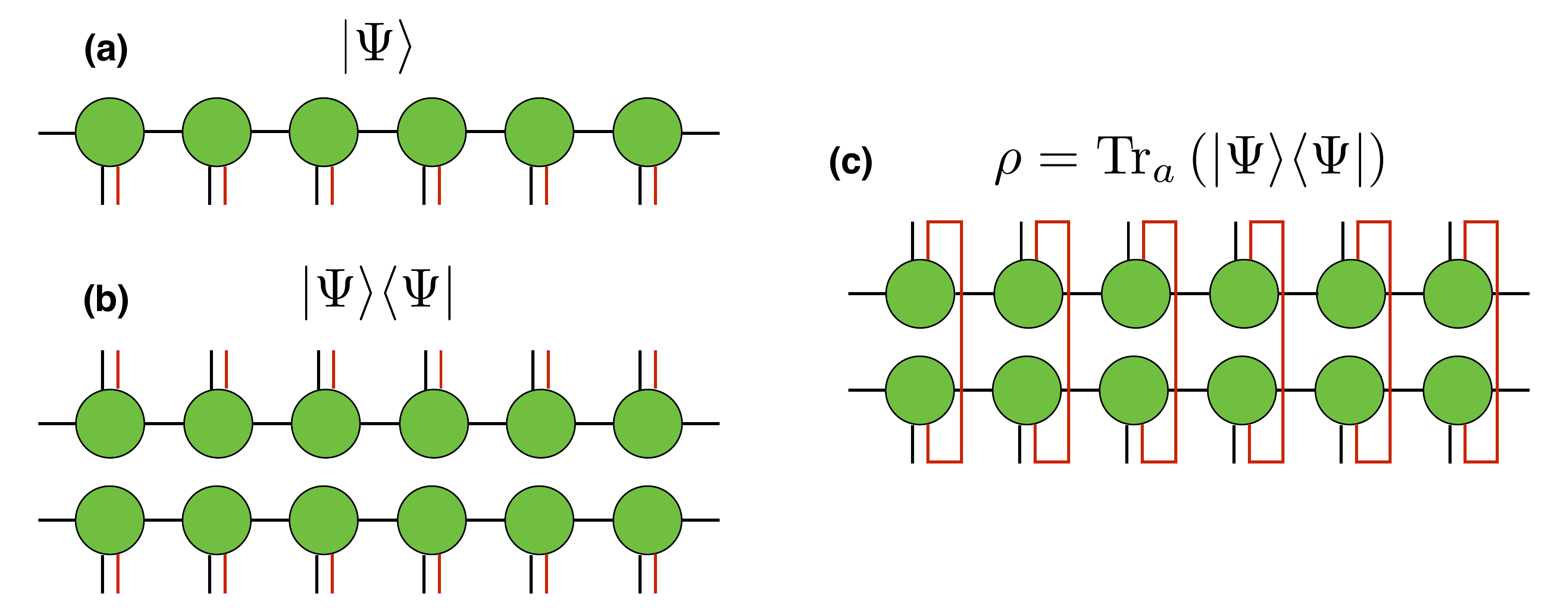}
				\caption{(a) Defining an MPS $|\Psi \rangle$ over the enlarged Hilbert space using ancilas (in red) (b) Taking the projector of the MPS with ancillas (c) Tracing out the ancillas from the projector to obtain the MPDO $\rho$.}
				\label{MPDO1}
			\end{center}
		\end{figure}

	In 2004, 
	\cite{VerstraeteMPDO} introduced the concept of MPDO which extended the MPS formalism from pure to mixed states. Let us recall that an MPS can be written in the following form
		\begin{equation}
		|\psi\rangle = \sum_{s_1,\dots,s_N=1}^{d} A_1^{s_1}\dots A_N^{s_N}|s_1, \dots, s_N \rangle
		\end{equation}
		where the $A$'s are matrices whose dimension is bounded by some fixed number $D$ (also called the bond dimension $\chi$) and $d$ is the physical dimension of the Hilbert space. An MPDO $\rho$ of $N$ $d$-level particles with $(D_1,D_2,\dots,D_N)$-dimensional bonds is then defined as
		\begin{multline}
		\rho = \sum_{s_1,s_1',\dots, s_N,s_N' = 1}^{d} (M_1^{s_1,s_1'}\dots M_N^{s_N,s_N'})|s_1,\dots,s_N \rangle\\ \times \langle s_1', \dots, s_N'|,
		\end{multline}
		where $M_k^{s_k,s_k'}$ are $D_k^2 \times D_{k+1}^2$ matrices that can be decomposed as
		\begin{equation}
		M_k^{s,s'}= \sum_{a=1}^{d_k} A_k^{s,a} \otimes (A_k^{s',a})^\ast. 
		\end{equation}
		where $d_k$ is at most $dD_kD_{k+1}$ and the matrices $A_k^{s,k}$ are of size $D_k \times D_{k+1}$. Such a construction of MPDOs automatically ensures the positivity of the reduced density matrix $\rho$. This is shown in Fig.~\ref{MPDO}. 
		This MPDO can be expressed in terms of a pure state MPS by defining it over a larger Hilbert space and using the concept of purification~\cite{Nielsen2000}. This can be done by associating an ancilla with a Hilbert space of dimension $d_k$ with each physical system. One can then choose an orthonormal basis $|s_k, a_k \rangle$ for these physical and ancilla indices. The corresponding MPS for this system can be written as
		\begin{equation}
		|\Psi \rangle = \sum_{s_1,\dots, s_N} \sum_{a_1,\dots, a_N} A_{1}^{s_1,a_1} \dots A_{N}^{s_N, a_N} |s_1a_1, \dots, s_Na_N \rangle
		\end{equation}
		The MPDO $\rho$ can be obtained by tracing over the ancillas i.e. $\rho = \text{Tr}_a\left( |\Psi \rangle \langle \Psi |\right)$. This process is illustrated in Fig.~\ref{MPDO1}.
		The original $A_k$ matrices can be recovered from $M_k$ by doing some eigenvalue decomposition. To determine the evolution of a Hamiltonian of a mixed state in real and imaginary time, they simply simulated the evolution of the purification by updating the $A_k$ matrices using an iterative procedure similar to the standard DMRG in this technique. The purification could then be used to reconstruct the density operator at any time and compute the expectation values of the observables. Such a purification scheme can be used for mixed state evolution under dissipation as well as for thermal equilibrium and can be implemented irrespective of periodic or open boundary conditions, finite or infinite systems. The main source of errors in this procedure, like most other TN techniques are (i) Trotter error and (ii) truncation error.
		Such an approach with ancillas was also applied to study the thermodynamic properties of several spin chains in Ref.~\cite{Whiteancilla}. Although the MPDOs in Ref.~\cite{VerstraeteMPDO} are positive by construction, it was shown in Ref.~\cite{purification_problem} that such a MPDO descriptions of mixed states are not exactly equivalent to the one obtained using local purification schemes. In particular, it was shown that the bond dimension of the locally purified MPS $D'$ is not upper bounded by the bond dimension of the MPDO $D$. In fact, the local purification techniques can be much more costly than the MPDO form itself. Thus, the authors concluded that a description of mixed states which is both efficient and locally positive semi definite does not exist and that one can only make approximations. 
		
\begin{figure}
			\begin{center}
				\includegraphics[scale = 0.24]{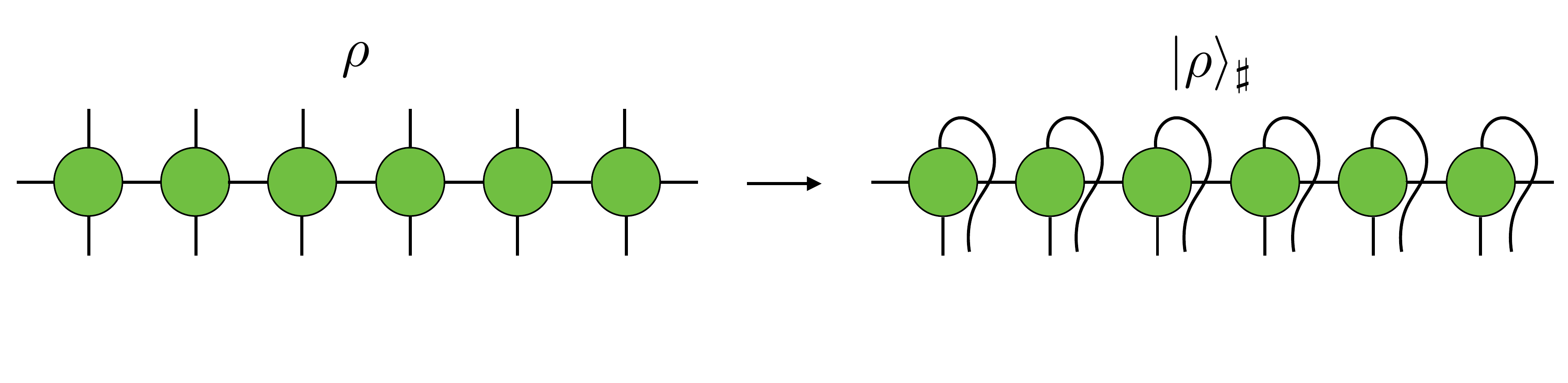}
				\caption{Choi isomorphism: vectorizing a density matrix written in terms of an MPO. In TN diagram, it is simply reshaping one of the indices and gluing it with the other thereby giving us an MPS.}
				\label{choi_DM}
			\end{center}
		\end{figure}

Around the same time, \cite{Zwolak2004} proposed another technique to study the mixed state dynamics in one dimensional lattice systems. Their technique, which is also based on MPS, used the Time Evolving Block Decimation (TEBD) to simulate the real time Markovian dynamics given by a master equation with nearest-neighbor couplings. At the heart of this algorithm, lies the concept of `Choi isomorphism'. It is more of a mathematical trick and it states that one can rewrite the coefficients of a matrix as those of a vector. In other words, this is simply turning a bra index into a ket index for a density matrix (understanding the coefficients of $\rho$ as those of a vectorized density matrix denoted by $|\rho\rangle_{\sharp}$). And in the language of TN diagrams, it can be regarded as reshaping one of the legs and gluing it with the other (Fig. \ref{choi_DM}).
		Once vectorized, $|\rho\rangle_{\sharp}$ now lives in the $n$-fold tensor product of $\mathbb{C}_{d^2}$ and the master equation can be written in the vector form.
		The mixed state will now look like as follows
		\begin{equation}
		|\rho\rangle_{\sharp} = \sum_{i_1 = 0}^{d^2-1} \cdots \sum_{i_N = 0}^{d^2-1} c_{i_1 \cdots i_N} |i_1\rangle_{\sharp} \otimes \cdots \otimes |i_N\rangle_{\sharp}.
		\end{equation}
		where $|i_l\rangle_{\sharp}$ is an orthonormal basis of $\mathbb{C}_{d^2}$ for site $l$. Further assuming that the Liouvillian superoperator $\mathcal{L}$ can be decomposed into terms involving at most nearest-neighbors i.e. $\mathcal{L}[\rho] = \sum_l \mathcal{L}_{l,l+1}[\rho]$, one could in principle use the usual TEBD algorithm to solve Eq.~\ref{master0} by starting from some initial Matrix Product Operator (MPO) (shown in in the left side of Fig. \ref{choi_DM}). This was the basic idea behind the technique in Ref.~\cite{Zwolak2004}. \ak{One of the first applications of this technique was the study of the driven-dissipative Bose-Hubbard model in the context of optical resonators \cite{Hartmann2010}.} More detailed explanation of this vectorization process will be explained later when we discuss the case for higher dimensional systems. Although the technique proved to be extremely simple and efficient, the issue of positivity still remained at large. In fact, checking the positivity of a reduced density matrix is known to be a very hard problem in physics~\cite{jenspositivity}.
	
        Another approach was taken in Ref.~\cite{Werner2016} to solve the problem of positivity. In this approach, instead of expressing $\rho$ directly as an MPO, at every stage of the algorithm, $\rho$ was kept in its locally purified $\rho = X X^{\dagger}$, where the purification operator $X$ is decomposed as a variational tensor network. 
		\begin{equation}
		[X]^{s_1,\dots,s_N}_{r_1,\dots, r_N} = \sum_{m_1,\dots,m_{N-1}}A^{[1]s_1,r_1}_{m_1} A^{[2]s_2,r_2}_{m_1,m_2}\dots A^{[N]s_N,r_N}_{m_{N-1}}
		\end{equation}
		where $1\leq s_l \leq d$ , $1\leq r_l \leq K$ and $1\leq m_l \leq D$. $A^{[l]}$ are rank-four tensors with physical dimension $d$, bond dimension $D$ and Kraus dimension $K$. Then, a technique similar to the usual TEBD was used to update the tensors. Such an approach never required to contract the two TN layers ($X$ and $X^\dagger$) together, thereby ensuring positivity at all times during the evolution. The technique also provided more control of the approximation error with respect to the trace norm.

 In Ref. \cite{Cui2015}, a very interesting and different approach based on MPO was taken for finding the steady states of dissipative 1D systems governed by the master equation of the Lindbladian form, $\frac{d \rho}{dt} = \mathcal{L}[\rho]$, where $\mathcal{L}$ is the Liouvillian superoperator. In this technique, instead of doing the full real time evolution of the Liouvillian, they proposed a variational method that searches for the null eigenvector of $\mathcal{L}$ which is, by definition, the steady state of the master equation in the Lindbladian form.  Their results were based on the principle that if $\rho_s$ is the steady state of the Lindbladian master equation satisfying $\hat{\mathcal{L}}|\rho_s \rangle_{\sharp}=0$, then $|\rho_s\rangle_{\sharp}$ will also be the ground state of the non-local Hamiltonian $\hat{\mathcal{L}}^\dagger \hat{\mathcal{L}}$ (since it is Hermitian and positive semi-definite) where $|\rho_s\rangle_{\sharp}$ is the vectorized form of the steady state density matrix. Then using a variational algorithm, they directly targeted the ground state of $\hat{\mathcal{L}}^\dagger \hat{\mathcal{L}}$ to find the steady state of the Lindbladian master equation for a finite chain. One of the reasons why directly targeting the ground state of $\hat{\mathcal{L}}^\dagger \hat{\mathcal{L}}$ might be advantageous is that unlike imaginary time evolution, where the sequence of states visited by the algorithm is unimportant, the simulation of a master equation requires us to follow real time evolution. Therefore, if there are errors in the intermediate states visited by the algorithm, it may lead to problems in the convergence of our steady state. For example, some of the intermediate states may require large bond dimensions of the MPO although it is known that the final steady state can be well-represented by an MPO of small bond dimensions~\cite{Cai2013,Bonnes2014}. Also, one doesn't need to worry about the large entanglement growth of real time evolution. A very similar approach was taken in Ref.~\cite{Mascarenhas2015} where the algorithm instead of doing a time-evolution, searched for the null eigenvalue of the Liouvillian superoperator $\mathcal{L}$ by sweeping along the system. Their method claimed to work even in the weakly dissipative regime by slowly tuning the dissipation rates along the sweeps. However, it needs to be noted that such techniques, while advantageous numerically, cannot be used for obtaining the transient states.
		
In another paper~\cite{Gangat2017}, this idea was applied to infinite 1D systems (i.e. the thermodynamic limit) using a hybrid technique of both imaginary and real time evolution. They took a local auxiliary Hamiltonian $\mathcal{H}$ whose ground state is a good approximation to the ground state of the nonlocal Hamiltonian $\hat{\mathcal{L}}^\dagger \hat{\mathcal{L}}$ by taking its $k^{\text{th}}$ root as 
		\begin{equation}
		\mathcal{H} = \sum_{r \in \mathbb{Z}}(\hat{\mathcal{L}}^\dagger_r \hat{\mathcal{L}_r})^{1/k}
		\end{equation}
		where $\hat{\mathcal{L}}=\sum_{r\in \mathbb{Z}}\hat{\mathcal{L}}_r$ since $\hat{\mathcal{L}}$ is a translationally invariant local operator. The $k^{\text{th}}$ root was taken in order to yield faster convergence. The idea is that if the gap between the two lowest eigenvalues of $\hat{\mathcal{L}}^\dagger_r \hat{\mathcal{L}_r}$ is less than one, then $k>1$ will increase the gap since $\hat{\mathcal{L}}^\dagger_r \hat{\mathcal{L}_r}$ is positive semi definite, thereby achieving faster convergence to the ground state. The authors then performed a real time evolution to obtain a more accurate steady state. In quick summary, the main steps of the algorithm is:
		\begin{enumerate}
		\item[(i)] Imaginary time evolution of the auxiliary Hamiltonian $\mathcal{H}$ starting from some vectorized initial density matrx $|\rho_0 \rangle$
		\begin{equation}
			|\rho_G \rangle \approx \lim_{\tau \rightarrow \infty} \frac{e^{-\mathcal{H}\tau}|\rho_0 \rangle}{||e^{-\mathcal{H}\tau}|\rho_0 \rangle ||}
		\end{equation}
		
		\item[(ii)] Real time evolution of the Liouvillian superoperator starting from $|\rho_G \rangle$
		\begin{equation}
			|\rho_S \rangle \approx \lim_{T \rightarrow \infty} \frac{e^{\mathcal{L}T}|\rho_G \rangle}{||e^{\mathcal{L}T}|\rho_G \rangle ||}
		\end{equation}
		\end{enumerate}
	where 	$|\rho_S \rangle$ is the desired steady state of the Liouvillian master equation. Imaginary time evolution in step(i) ensures that one does not pass through highly entangled transient regime. Step(ii) increases the accuracy of the stationary state since $|\rho_G \rangle$ is the ground state of $\mathcal{H}$ which is a truncated approximation of  the non local Hamiltonian $\hat{\mathcal{L}}^\dagger \hat{\mathcal{L}}$.
	
  It is worth mentioning that in one spatial dimension, many of the above techniques and their combinations have been used for studying not only other important disspative models \cite{Pizorn2013,Honing2012,Mascarenhas2015,Carollo2019}, \hw{including the dissipative Ising model of Eq.~(\ref{eq:dising}) \cite{Honing2013,Mendoza-Arenas2016},}
  but also in dissipative preparation of topologically ordered materials \cite{Iemini2016} \ak{as well as in the energy transport \cite{Guotransport}}. \ak{Very recently, MPO based techniques have been applied to study vibronic states which extends the application to quantum biology and organic photo voltaics \cite{DAMPF}. 
  and also to study the dynamics of photonic circuits with time delays and quantum feedback \cite{Zoller_photoniccircuit}. We do not discuss the later two works due to the non-Markovian nature of the problem which is beyond the scope of this review. Similar MPS based techniques that go beyond the Lindblad master equation \cite{Guoredfield} or the markovian approximation \cite{Guonon-markov} are also not discussed here.}
	
\subsection{Extensions to higher dimensions}

\label{sec:tn2d}

	Unlike the case for 1d, the generalization of MPS in higher dimensions, also known as Projected Entangled Pair States (PEPS) \ak{or Tensor Product States (TPS)} comes with some serious limitations and there are still many open problems \cite{Ciracpepsopenproblem}. Not only do the PEPS algorithm require serious programming effort, exact contratcion of PEPS is known to be a mathematically hard problem \cite{Schuchpepshard,Eisertpepshard}. To achieve this, one requires additional PEPS contraction algorithms \cite{Romanctm1,Romanctm2,bMPS} that are nevertheless known to give very accurate results, \ak{in particular, for gapped systems. Even for critical systems with algebraically decaying correlations, the PEPS contraction schemes are known to provide reasonably accurate results with sufficiently high bond dimension of the environment \cite{Romanctm2}. In fact, recently, techniques have been introduced to capture the infinite correlation length of 2D critical systems using iPEPS based on finite correlation length scaling \cite{Corbozgaplessipeps,Lauchligaplessipeps}}. Thus, despite the higher requirement of numerical dedications and limitations, PEPS algorithms are becoming state of the art numerical tools for strongly correlated two dimensional systems. Recently, PEPS provided the best variational energy for the 2D Hubbard model \cite{CorbozHubbard}, have offered several new insights on paradigmatic models and real materials in the lab \cite{Xiangkhaf,CorbozSsland,CorbozSsland2,Kshetrimayummaterial}. 
	The successes of PEPS so far, however, is mostly confined to ground state calculations and partially to thermal states \cite{PiotrThermal1,PiotrThermal2,PiotrThermal3,Kshetrimayumthermal,Zhouthermal2D} \ak{using the concept of Projected Entangled Pair Operators (PEPOs) or Tensor Product Operators (TPOs), which we will discuss in more detail later, and, more recently, to time evolution \cite{Claudiusevolution,Piotrpepstimeevolution,Kshetrimayumevolution}}. For the context of open \ak{dissipative} quantum system, so far there is only one known approach using PEPS \cite{Kshetrimayum2017} and another one using a Corner Space Renormalization method \cite{Finazzi2015}. We describe them below. We will also discuss briefly other potential implementation techniques and possible issues while using PEPS formalism in particular for such open systems.

\begin{figure}
			\begin{center}
				\includegraphics[scale = 0.3]{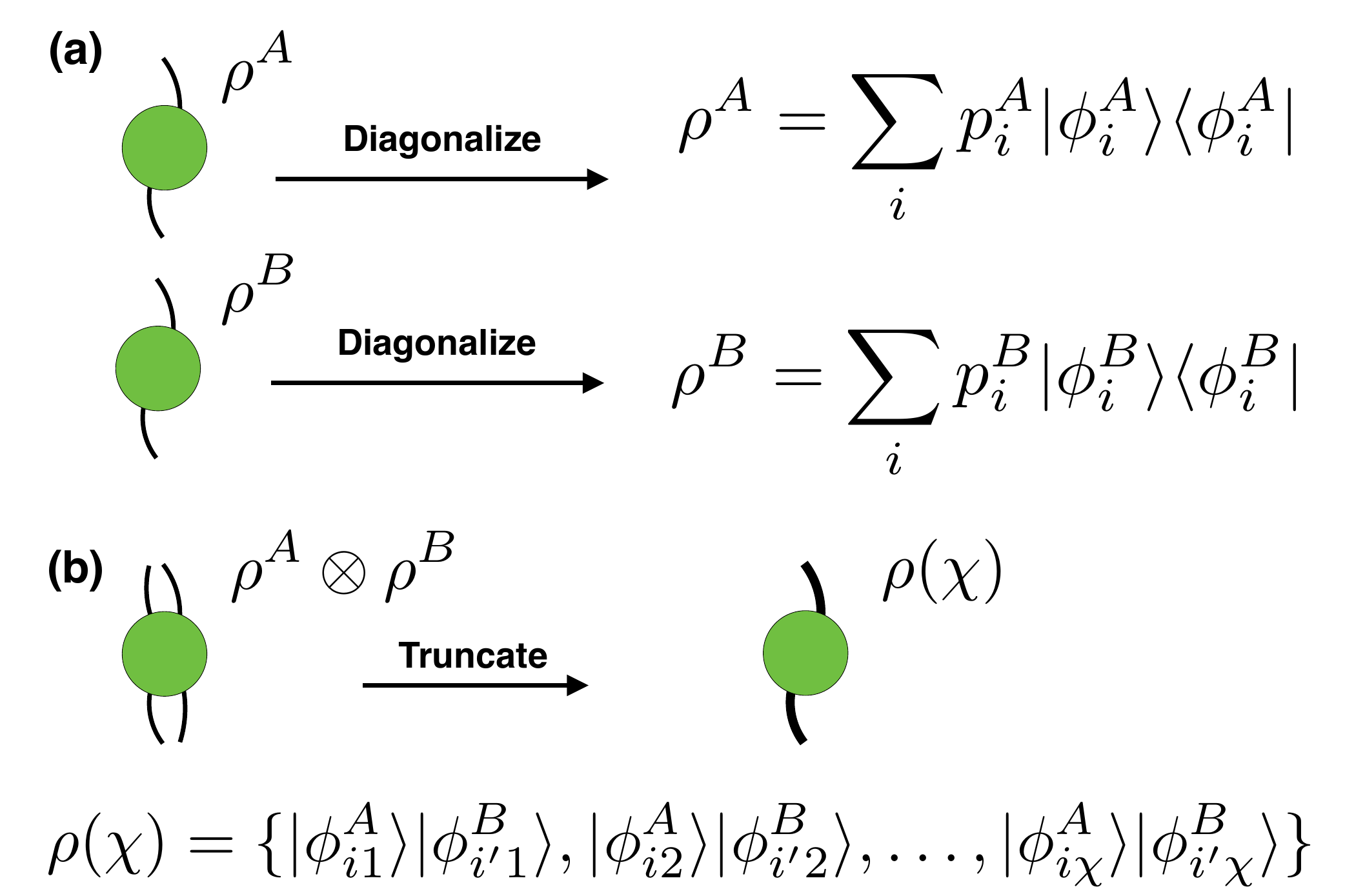}
				\caption{(a) Steady state density matrices of two systems $A$ and $B$ are first obtained using brute force. They are then expressed in their respective diagonal forms. (b) We then merge the two systems and keep only the $\chi$ most probable pair of states. The process is repeated for different $\chi$s until we get some convergence. Larger systems can be simulated by merging more systems in step (b). }
				\label{CSR}
			\end{center}
		\end{figure}

 The Corner Space Renormalization method~\cite{Finazzi2015} solves the master equation in a corner of the Hilbert space through an iterative procedure. It starts by finding the steady state density matrix for small lattice systems (say $\rho^A$ and $\rho^B$ for systems $A$ and $B$ respectively). This can be done by a brute force integration of the master equation since the system size is very small. The steady state density matrices can be diagonalized and written as
		\begin{equation}
		\begin{split}
		\rho^A = \sum_i p_i^A |\phi_i^A \rangle \langle \phi_i^A|,\\
		\rho^B = \sum_i p_i^B |\phi_i^B \rangle \langle \phi_i^B|,
		\end{split}
		\end{equation}
		where the states $|\phi_i^A \rangle $ form an orthonormal basis for $\mathcal{H}_A$ (the Hilbert space corresponding to system $A$) and $p_i^A$ are the corresponding probabilities. Similar notation follows for system $B$. The two systems are then merged and the $\chi$ most probable product states spanning the so-called corner space are selected i.e. we only keep the subspace generated by the orthonormal basis $\{ |\phi_{i1}^A\rangle |\phi_{i'1}^B\rangle, |\phi_{i2}^A\rangle |\phi_{i'2}^B\rangle, \dots, |\phi_{i{\chi}}^A\rangle |\phi_{i'{\chi}}^B\rangle  \}$ where the product of the probabilities of the two systems are arranged in decreasing order of magnitude. In this way, we only keep the $\chi$ most probable pair of states. The steady state of the density matrix in this corner space can be determined either by direct numerical integration in time (for small $\chi$) or by using a stochastic wave function Monte Carlo algorithms for large $\chi$, see Sec.~\ref{sec:wfmc}. One can then increase the size of the corner $\chi$ until convergence in some observables is reached. Larger systems can be simulated by merging more systems as we discussed in the initial steps. A simplified summary of the steps involved is shown in Fig.~\ref{CSR}. 

The proposed CSR method was used to study the driven-dissipative Bose-Hubbard model in 2D in both periodic and open boundary conditions for system sizes up to \ak{$16\times 16$ lattice sites. The technique has also been used to study the critical Heisenberg model \cite{Ciuticritical} for system size up to $6\times6$ lattice sites and more recently the critical regime in the Bose Hubbard model \cite{SavonnaCSR} for up to $8\times8$ lattices. The size of the lattice that can be simulated using this technique depends on the entanglement of the steady state.} 
Even if not obvious at first sight, the structure of the density operator generated by the Corner Space Renormalization method amounts to that of a Tree Tensor Network \cite{TTNvidal}. As such, this particular method, even if understood in terms of TNs, is tailored to driven-dissipative systems of finite size. For generalizing it to the thermodynamic limit or for non-driven non-dissipative systems, one needs to use more general TN techniques. We will discuss one such technique which we developed recently below.

\begin{figure}
	\begin{center}
		\includegraphics[width=0.5\textwidth]{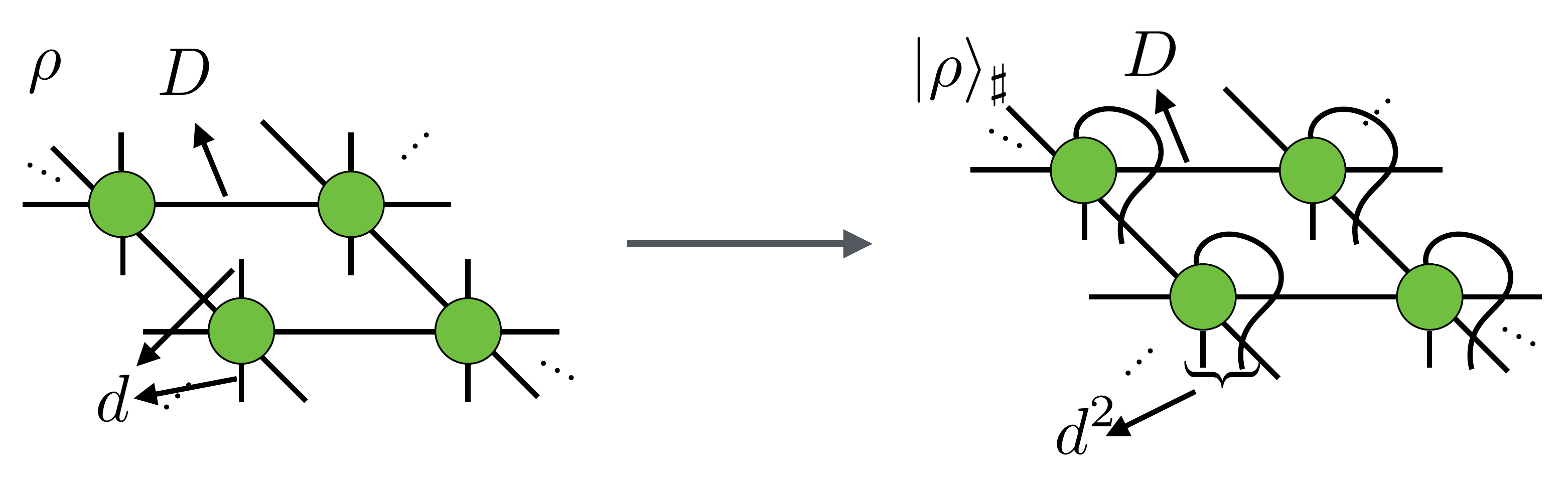}
		\caption{TN diagram for the PEPO of $\rho$ on a 2d square lattice, with bond dimension $D$ and physical dimension $d$. When vectorized, it can be understood as a PEPS for $|\rho\rangle_\sharp$ with physical dimension $d^2$}
		\label{pepo2peps}
	\end{center}
\end{figure}

In \cite{Kshetrimayum2017}, we make use of the concept of PEPO by vectorizing them. \ak{PEPOs} are simply the operator version of \ak{PEPS}, in the same way that an \ak{MPO} is the operator version of \ak{MPS} for the 1d case. Hence, PEPOs are used to represent mixed states $\rho$ in 2D, even beyond dissipative systems, e.g., for thermal states \cite{PiotrThermal1,Kshetrimayumthermal,PiotrThermal2,PiotrThermal3}.
As mentioned before, such a construction of density matrices using PEPOs does not automatically guarantee the positivity of the density matrix. However, for simulations targeting the steady states, this lack of exact positivity is not a bottleneck if the fixed point is not very highly entangled. For the moment, we will restrict our discussion to this case. Once we have our PEPO, we vectorize it i.e. rewrite the coefficients of the PEPO as a PEPS (also called Choi's isomorphism). Once vectorized, the PEPO $\rho$ can be understood as a \ak{PEPS} of physical dimension $d^2$ and bond dimension $D$ (now called $|\rho\rangle_{\sharp} $), as shown also in Fig. \ref{pepo2peps}. 
The vectorized form of the Lindblad master equation Eq.~(\ref{master0}) can be written as
\begin{equation}
\frac{d}{dt}|\rho\rangle_{\sharp} = \mathcal{L}_{\sharp}|\rho\rangle_{\sharp}
\label{lindveceq}
\end{equation}
where the vectorized Liouvillian operator is given by 
\begin{equation}
\begin{split}
\mathcal{L}_\sharp& \equiv -i \left(H\otimes \mathbb{I} - \mathbb{I}\otimes H^T \right)\\
&+ \sum_\mu \left(L_\mu\otimes L^*_\mu -\frac{1}{2}L_\mu^\dagger L_\mu \otimes\mathbb{I} - \frac{1}{2}\mathbb{I}\otimes L^*_\mu L_\mu^T \right). 
\end{split}
\end{equation}
$H$ is the Hamiltonian of the system and $\mathbb{I}$ corresond to the identity operator. $L_{\mu}$ and $L_{\mu}^{\dagger}$ correspond to the on-site Lindblad/jump operators, responsible for dissipation. The tensor product $\otimes$ separates the operator acting on the ket and bra index of $\rho$ before the vectorization. When the vectorized Liouvillian superoperator $\mathcal{L}_\sharp$ is independent of time, Eq. \ref{lindveceq} can be integrated as 
\begin{equation}
|\rho(t)\rangle_\sharp = e^{\mathcal{L}_\sharp t} |\rho (0)\rangle_\sharp, 
\label{sol}
\end{equation}
where $|\rho(0)\rangle_{\sharp}$ is some vectorized initial density matrix, written as a PEPS. In the limit of $t\rightarrow \infty$, we obtain the non-equilibirum steady state (NESS) as the fixed point of the master equation which we denote by $|\rho_s\rangle_{\sharp}$. From Eq. \ref{lindveceq}, it is also obvious that $|\rho_s\rangle_{\sharp}$ is the right eigen vector of $\mathcal{L}_{\sharp}$ with zero eigen value so that
\begin{equation} 
\mathcal{L}_\sharp  |\rho_s\rangle_\sharp = 0 . 
\end{equation}

For a Liouvillian $\mathcal{L}$ consisting of local terms say $\mathcal{L}[\rho] = \sum_{\langle i, j \rangle}\mathcal{L}^{[i,j]}[\rho]$, the vectorized form of the Lindblad equation Eq.~(\ref{master0}) yields a parallelism with the calculation of ground states of local Hamiltonians by imaginary-time evolution, which we detail in Table \ref{tab1}. 
		\begin{table}[h]
			\centering
			\begin{tabular}{||c|c||} 
				\hline 
				~~~~~~Ground states~~~~~~ & ~~~~~~Steady states~~~~ \\
				\hline 
				\hline 			$H = \sum_{\langle i,j \rangle} h^{[i,j]}$ & $\mathcal{L}_\sharp = \sum_{\langle i, j \rangle}\mathcal{L}_\sharp^{[i,j]}$ \\
				$e^{-H t}$ &  $ e^{ \mathcal{L}_\sharp t}$ \\ 
				$|e_0\rangle$ & $|\rho_s\rangle_\sharp$ \\ 
				$\langle e_0| H |e_0\rangle = e_0$ & $_\sharp \langle|\rho_s\rangle \mathcal{L}_\sharp |\rho_s\rangle_\sharp = 0$ \\
				Imaginary time & Real time \\
				\hline
			\end{tabular}
			\caption{Ground state calculation in a closed quantum system (left) and Steady state calculation in an open quantum system (right). The former one requires an imaginary time evolution while the latter follows a real time evolution. Both the Hamiltonian $H$ and the vectorized Liouvillian $\mathcal{L}_{\sharp}$ can be decomposed as a sum of local terms. $|e_0\rangle$ is the ground state of the many-body Hamiltonian with $e_0$ as its ground state. $|\rho_s\rangle_{\sharp}$ is the non-equilibrium steady state of the Liouvillian in their vectorized forms.}
			\label{tab1}
		\end{table}
		Given the parallelism above, it is clear that one can adapt, at least in principle, the methods to compute imaginary time evolution of a pure state as generated by local Hamiltonians, to compute also the real time evolution of a mixed state as generated by local Liouvillians. This was, in fact, the approach taken in Ref.\cite{Zwolak2004} for finite-size 1d systems, using Matrix Product Operators (MPO) to describe the 1d reduced density matrix, and proceeding as in the Time-Evolving Block Decimation (TEBD) algorithm for ground states of 1d local Hamiltonians \cite{tebd1,tebd2,tebd3,tebd4} as we have discussed previously. In \cite{Kshetrimayum2017}, we extended this implementation for the case of 2D systems using the concept of  \ak{PEPO} with physical dimension $d$ and bond dimension $D$, see Fig.\ref{pepo2peps}. 
For the case of an infinite-size 2d system, this setting is actually equivalent to that of the infinite-PEPS algorithm (iPEPS) to compute ground states of local Hamiltonians in 2d in the thermodynamic limit. Thus, in principle, one can use the full machinery of iPEPS to tackle as well the problem of 2d dissipation and steady states.

\begin{figure}
	\begin{center}
		\includegraphics[width=0.36\textwidth]{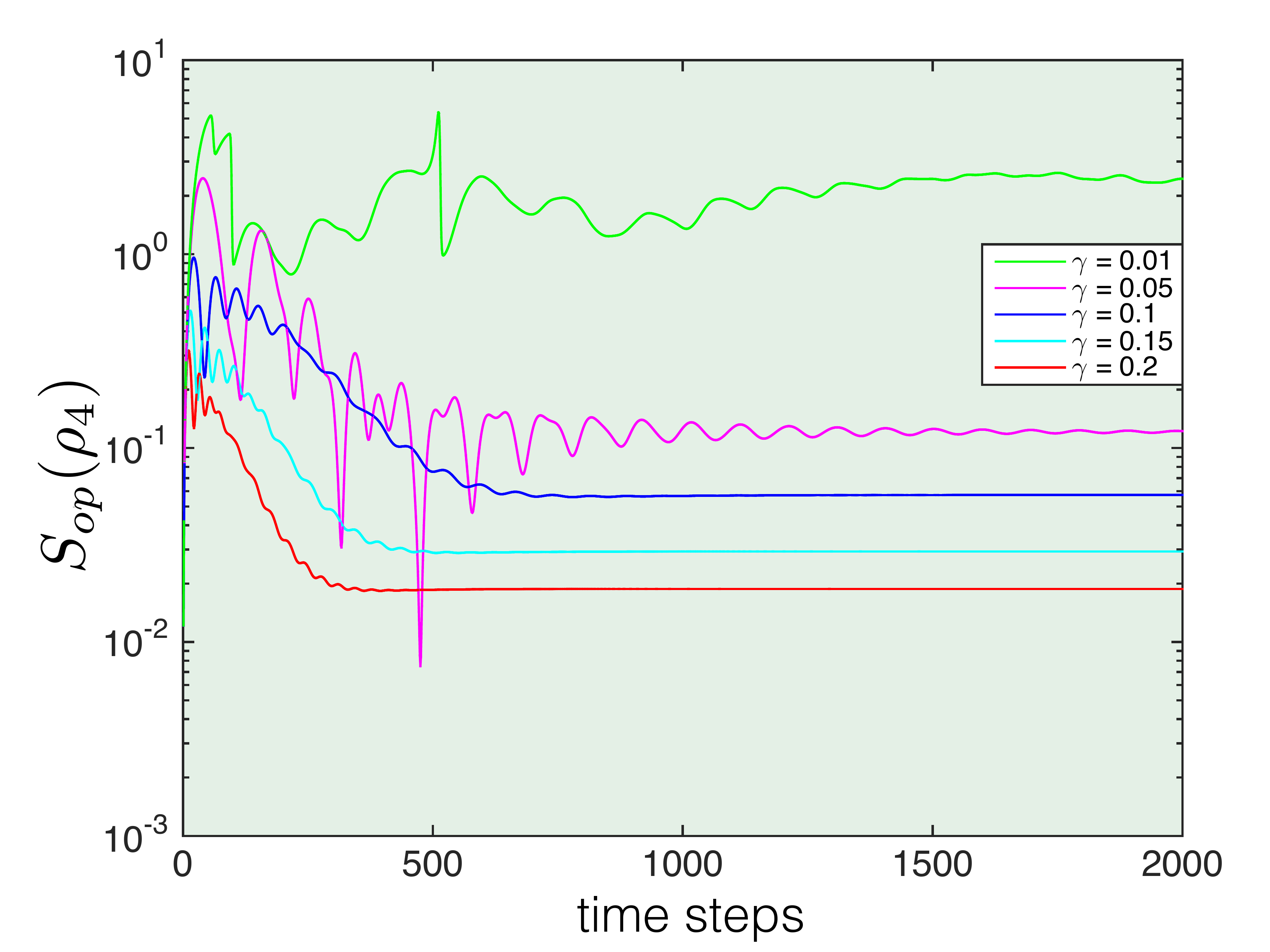}
		\caption{Operator entanglement entropy $S_{op}$ for a block of $2 \times 2$ unit cell for real time evolution of the master equation for different values of dissipation strength. Stronger dissipation implies lower entanglement growth and faster convergence to the NESS. Figure taken from \cite{Kshetrimayum2017} for the dissipative Ising model.}
		\label{pepoentropy}
	\end{center}
\end{figure}

There seems to be, however, one problem with this idea: unlike in imaginary-time evolution, we are now dealing with real time. In the master equation, part of the evolution is generated by a Hamiltonian $H$, and part by the Lindblad  operators $L_\mu$. The Hamiltonian part corresponds actually to a unitary ``Schr\"odinger-like" evolution in real time, which typically increases the   ``operator-entanglement" in $|\rho\rangle_\sharp$, up to a point where it may be too large to handle for a TN representation (e.g., 1d MPO or 2d PEPO) with a reasonable bond dimension. In 1d this is the reason why the simulations of master equations are only valid for a finite amount of time. In 2d, simple numerical experiments indicate that in a typical simulation the growth of entanglement is even faster than in 1d. Luckily, this is not a dead-end: if the dissipation is strong compared to the rate of entanglement growth, then the evolution drives the system into the steady state before hitting a large-entanglement region. In fact, even if there is too much entanglement for the TN at intermediate times, the dissipation may still drive the evolution towards a good approximation of the correct steady state. In short, dissipation limits the growth of entanglement if the fixed point attractor is strong enough. This can be verified numerically by plotting the operator entanglement entropy for different dissipation strengths as it flows into the NESS. This is shown in Fig. \ref{pepoentropy}. Details on how to compute this quantity can be found in \cite{Kshetrimayum2017,Kshetrimayumthesis}.

\begin{figure}
	\begin{center}
		\includegraphics[width=0.4\textwidth]{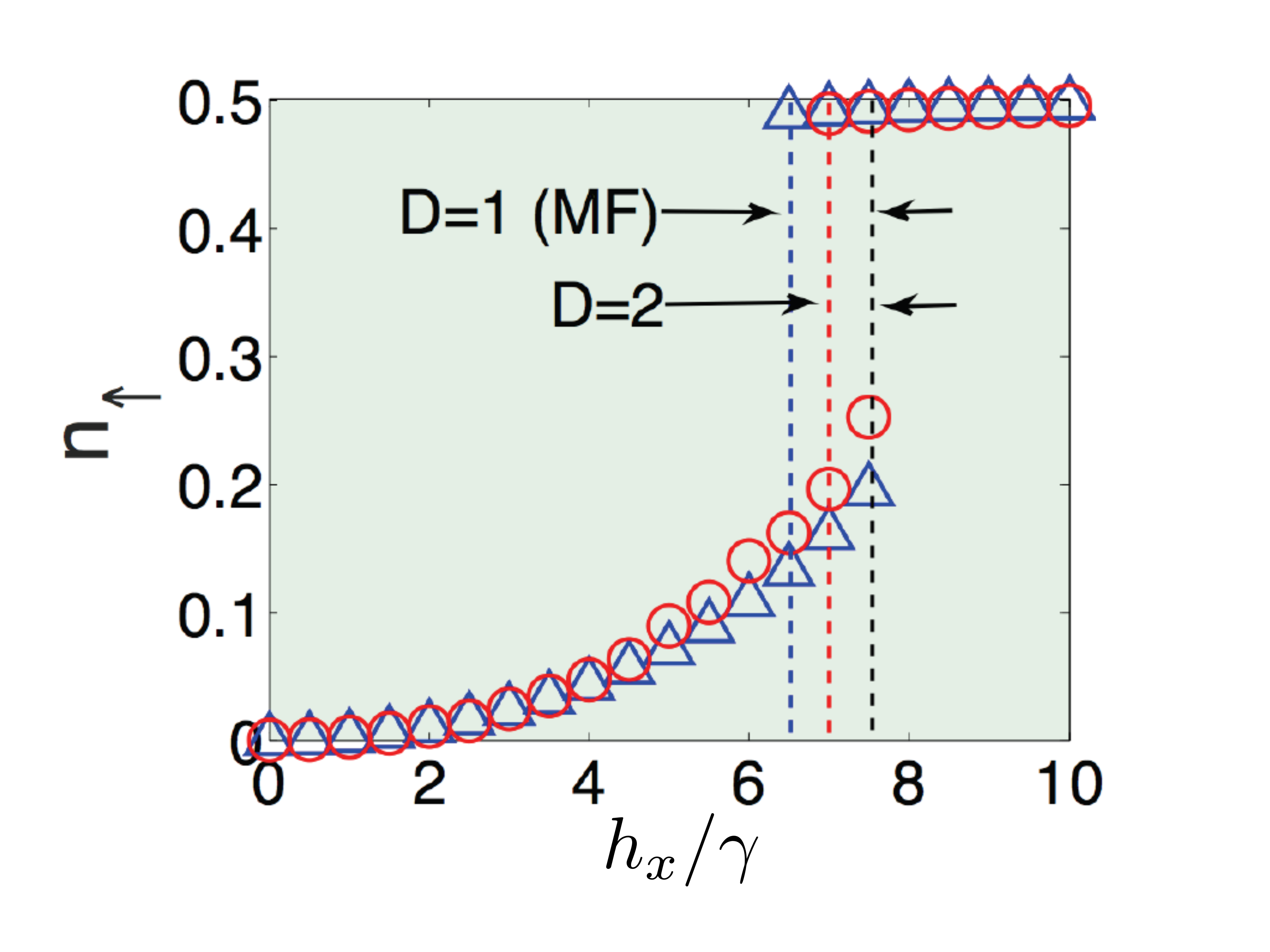}
		\caption{Our study, based on iPEPO found bistability in the phase diagram of the dissipative Ising model for low bond dimensions $D=1,2$. The bistability is replaced by a first order transition for higher $D$s. Figure taken from \cite{Kshetrimayum2017} for the dissipative Ising model.}
		\label{pepobistability}
	\end{center}
\end{figure}

Hence, one can apply the iPEPS machinery to compute the time evolution in 2d with a local Liouvillian  $\mathcal{L}$ and some initial state. This procedure was used to investigate the dissipative Ising and the XYZ model, confirming and offering several insights that were inaccesible before using mean field and other techniques. For example, for the dissipative Ising model of Eq.~(\ref{eq:dising}), given by the Hamiltonian $H = \frac{V}{4} \sum_{\langle i,j \rangle} \sigma_z^{[i]} \sigma_z^{[j]} + \frac{h_x}{2} \sum_i \sigma_x^{[i]}$ and Lindblad operators $L_\mu = \sqrt{\gamma} \sigma_-^{[\mu]}$. The phase diagram is controversial with some papers suggesting the existence of a bistable steady state \cite{Lee2011,Marcuzzi2014} and others supporting a first order transition \cite{Weimer2015a,Weimer2015,Maghrebi2016}. Our technique found bistability for low bond dimensions of the PEPO ($D=1,2$) which was replaced by a first order transition for higher $D$s, thus confirming that the bistability is an artifact of mean field. This is shown in Figure \ref{pepobistability}.
Furthermore, some studies suggested the existence of an antiferromagnetic region in the presence of the transverse field $h_x$ \cite{Lee2011,Weimer2015a}. Once again, while our technique found evidence of such an AF region, it eventually shrank with increasing bond dimension until it finally disappears for large enough $D$. Results of the dissipative Ising model have been reproduced independently using a different update scheme \cite{Piotrpepstimeevolution} compared to the one used in \cite{Kshetrimayum2017}. While the technique also employed vectorization along with iPEPS, the update scheme is based on maximizing the fidelity between two consecutive steps of the update of the iPEPS tensors. For the case of the dissipative Heisenberg model \cite{Lee2013} with the Hamiltonian
\begin{equation}
  H = \sum\limits_{\langle i,j \rangle} \left( J_x \sigma_x^i \sigma_x^j + J_y \sigma_y^i \sigma_y^j + J_z \sigma_z^i \sigma_z^j \right)
  \label{eq:dheisen}
\end{equation}
and the same Lindblad operators as before, our studies found no phenomenon of re-emergence in the phase diagram, confirming a prediction by studies using cluster mean-field approaches \cite{Jin2016}.
		
\begin{figure}
	\begin{center}
		\includegraphics[width=0.4\textwidth]{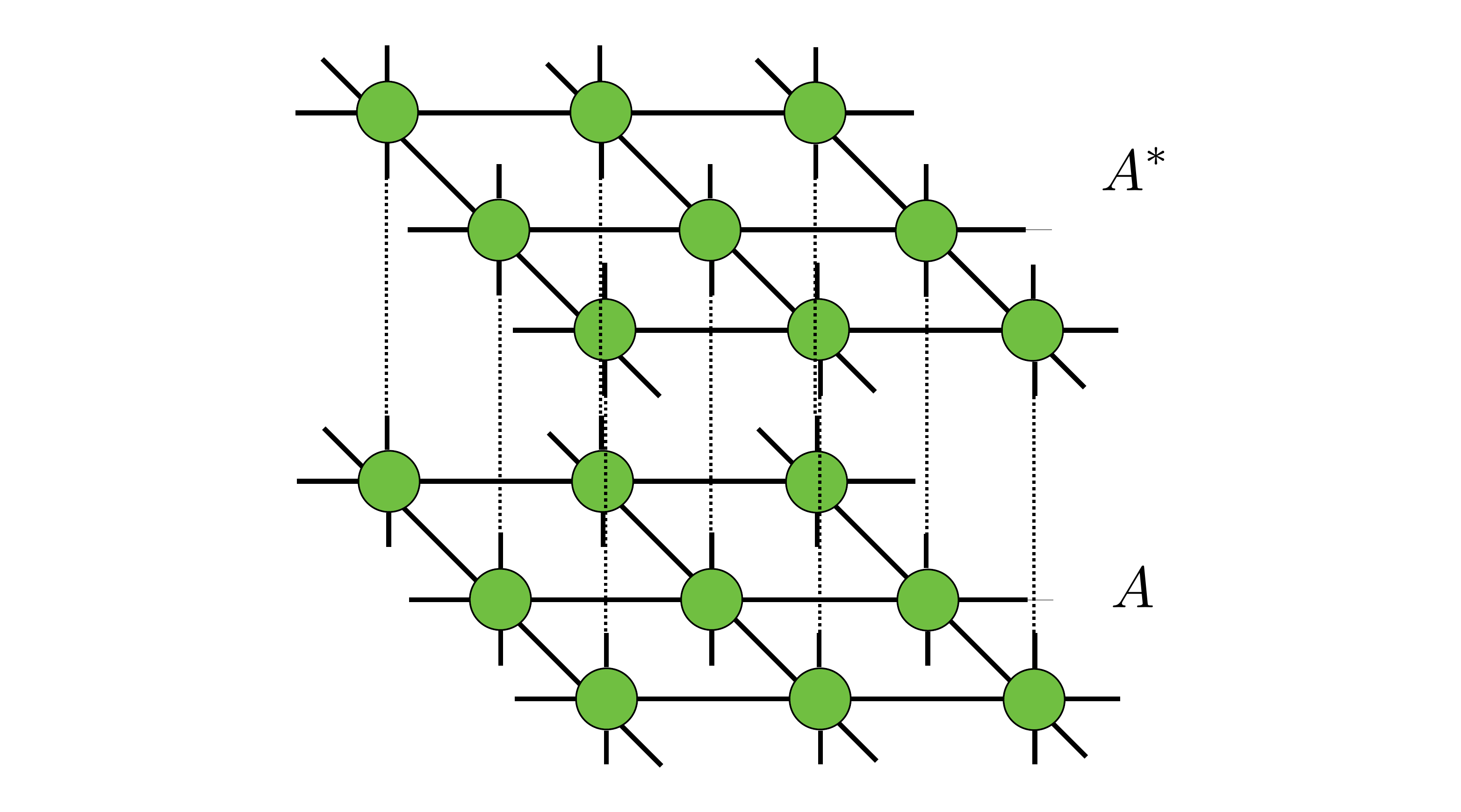}
		\caption{TN diagram for the PEPDO of $\rho$ on a 2d square lattice, with bond dimension $D$ and physical dimension $d$. When vectorized, it can be understood as a PEPS for $|\rho\rangle_\sharp$ with physical dimension $d^2$}
		\label{pepdo}
	\end{center}
\end{figure}

To the best of our knowledge, we have discussed most of the state-of-the-art numerical techniques based on TN for the study of open quantum many-body systems in both one and two spatial dimensions. We would, now, like to discuss some of the possible ideas that could be helpful in improving the existing algorithm and possible new implementation techniques specially in 2d. First of all, we remark that the 2d algorithm suggested above does not guarantee the positivity of the density matrices. This problem can be solved by starting from an initial state that is positive by construction, for example taking the product of two PEPOs which are the conjugate of each other ($A$ and $A^*$). One can then think about using a positivity preserving algorithm such as the one in \cite{Werner2016}. Such an algorithm will ensure the positivity of the density matrix at all times of the evolution. We can call this initial density matrix as Projected Entangled Pair Density Operator (PEPDO) as shown in Figure \ref{pepdo}. 
While such an approach may avoid the problem of negative eigenvalues of the density matrix, in practice, it may require very high bond dimension of the PEPDO and one, therefore, needs to consider the practical aspect of the implementation.

The other possibility would be to target the ground state of the Hermitian and positive semidefinite operator $\mathcal{L}_\sharp^\dagger \mathcal{L}_\sharp$. This ground state could be computed, e.g., by an imaginary time evolution. However, there are two major hurdles associated with this approach. First, the crossed products in  $\mathcal{L}_\sharp^\dagger \mathcal{L}_\sharp$ are non-local, and therefore the usual algorithms for time evolution are difficult to implement unless one introduces extra approximations in the range of the crossed terms. Another option is to approximate the ground state variationally, e.g., via the Density Matrix Renormalization Group \cite{dmrg1,dmrg2} in 1d, or variational PEPS in 2d \cite{Corbozvariationalpeps}. In the thermodynamic limit, however, this approach does not look very promising because of the non-locality of $\mathcal{L}_\sharp^\dagger \mathcal{L}_\sharp$ mentioned before. In any case, one could always represent this operator as a PEPO (in 2d), which would simplify some of the calculations, but at the cost of introducing a very large bond dimension in the representation of $\mathcal{L}_\sharp^\dagger \mathcal{L}_\sharp$. For instance, if a typical PEPO bond dimension for $\mathcal{L}_\sharp$ is $\sim 4$, then for $\mathcal{L}_\sharp^\dagger \mathcal{L}_\sharp$ it is $\sim 16$, which in 2d implies  \emph{extremely} slow calculations. Another option would be to target the variational minimization of the real part for the expectation value of $\mathcal{L}$. 
	
\section{Variational methods}

Variational techniques are often very powerful tools to analyze
quantum many-body systems, as demonstrated by the successes of density
functional theory \cite{Kohn1999} and matrix product state approaches
\cite{Schollwock2011} for ground state problems. As we will discuss in
this section, variational methods can also be successfully applied to
open quantum many-body systems.
	
\subsection{The variational principle for open quantum systems}

Variational methods generically consist of two steps. The first step
is a paramterization of the state of the system in terms of a set of
variational parameters $\{\alpha_i\}$. For open quantum system, it is
convenient to parameterize the density matrix, i.e., $\rho =
\rho(\{\alpha_i\})$, although parameterizations based on statistical
ensembles of pure states are also possible \cite{Transchel2014}. The
second step is to identify a suitable functional that can be optimized
by tuning the variational parameters. For open quantum systems, it is
very natural to apply a variational principle to find the steady state
of the quantum master equation, which can be found by solving the
equation $\dot\rho=0$. Importantly, the exact steady state can no
longer be determined after the variational parametrization, as the
steady state will generically lie outside the variational
manifold. Hence, the best possible option is to find the variational
parameters that will minimize the functional $||\mathcal{L}\rho||$ for
a suitable norm \cite{Weimer2015}.

The correct norm for the variational optimization can be identified as
the trace norm $||\mathcal{L}\rho||= \trtxt{|\dot\rho|}$, i.e., the
sum of the absolute values of the eigenvalues of $\dot\rho$
\cite{Weimer2015}. This choice can be motivated on different
grounds. First, the trace distance being the natural distance measure
for density matrices \cite{Nielsen2000} is highly suggestive of the
trace norm being the natural norm for the tangent space
$\dot\rho$. This can be formalized in the sense that the trace norm
describes an optimal measurement to distinguish $\dot\rho$ from the
zero matrix \cite{Gilchrist2005}. A second way to motivate the trace
norm is to consider classes of possible alternatives. It can be shown
that all Schatten $p$-norms of the form $(|\dot\rho|^p)^{1/p}$ are
inherently biased towards the maximally mixed state for all values of
$p>1$ \cite{Weimer2015}. Since functionals with $p<1$ do not
constitute proper norms, this leaves the trace norm as the only valid
choice. \hw{One can also understand the variational principle as a
  direct solution of the overdetermined steady state equation
  $\mathcal{L}\rho=0$ in terms of a trace norm minimization.}

In general, the evaluation of the variational functional is still an
exponentially hard problem, as the computation of the trace norm
requires the diagonalization of the matrix $\dot\rho$. However, it is
possible to construct upper bounds to the variational norm that retain
the variational character \cite{Weimer2015} \hw{and appear to
  introduce only small quantitative deviations even close to phase
  transitions \cite{Weimer2015a}. The} upper bound depends
on the variational manifold and its tangent space, i.e., the degree of
additional correlations that can be build up by applying the
Liouvillian to states within the variational manifold. For example,
for a variational class of product states of the form
$\rho=\prod_i\rho_i$, the upper bound $D$ can be given as
\begin{equation}
  D = \sum\limits_{ij \in \mathcal{T}} \tr{|\dot\rho_{ij}|},
  \label{eq:D}
\end{equation}
where $\mathcal{T}$ contains pairs of sites that are connected to each
other by the Liouvillian \cite{Weimer2015}.

The variational principle has been applied to find the steady states
of the dissipative Ising and Bose-Hubbard models introduced in
Eqs.~(\ref{eq:dising}) and (\ref{eq:dhubbard}), respectively
\cite{Weimer2015,Weimer2015a}, as well as dissipative Ising models
including a $\mathbb{Z}_2$ symmetry \cite{Overbeck2017}, purely
dissipative Heisenberg models \cite{Weimer2017}, dissipative Rydberg
gases \cite{Weimer2015a}, dissipative ensembles of nitrogen-vacancy
centers \cite{Raghunandan2018}, entanglement generation in cavity QED
arrays \cite{Lammers2016}, and dissipative Fermi-Hubbard models
\cite{Kaczmarczyk2016}. In the latter case, the study of fermionic
models was realized by employing a two-dimensional Jordan-Wigner
transformation, where the appearance of nonlocal Wigner strings was
ruled out by the choice of the variational manifold.

In the case where the steady state of the system is close to
criticality, it is possible to construct a dissipative Ginzburg-Landau
theory based on the variational principle \cite{Overbeck2017}. The
essential step is to perform a series expansion of the variational
norm of Eq.~\ref{eq:D} in terms of an order parameter field $\phi(x)$
and its spatial gradient $\nabla \phi(x)$, leading to
\begin{equation}
  D[\phi] = \int dx\, \sum\limits_m v_m [\nabla \phi(x)]^m + \sum\limits_n u_n [\phi(x)]^n.
  \label{eq:gl}
\end{equation}
All the coefficients $v_n$ and $u_n$ can be calculated from the
microscopic quantum master equation. The series can be truncated at
low orders of $m$ and $n$, as higher order terms are irrelevant close
to criticality. In the case of steady states with thermal statistics
due to the presence of a dynamical symmetry \cite{Sieberer2013}, it is
possible to construct a Ginzburg-Landau-Wilson functional integral for
an effective partition function \cite{Hohenberg2015}, given by
\begin{equation}
  Z_\text{eff} = \int \mathcal{D}\phi \exp\left(-\beta_\text{eff} D[\phi]\right).
  \label{eq:field}
\end{equation}
Here, the effective inverse temperature $\beta_\text{eff}$ can be
derived from the $u_0$ coeffecient, as this coefficient captures the
strength of fluctuations beyond a spatially homogeneous order
parameter field \cite{Overbeck2017}. The subsequent statistical field
theory of Eq.~(\ref{eq:field}) can then be analyzed using standard
techniques such as the perturbative renormalization group.

Finally, the variational principle can also be extended towards the
full time evolution of open quantum systems \cite{Overbeck2016},
following very similar ideas discussed in the context of the
time-dependent variational principle \cite{Kraus2012,Transchel2014}. There, the variational functional is replaced by a variational integration of the quantum master equation for small time steps $\tau$. For example, in the lowest order Euler approximation, it is given by
\begin{equation}
  D = \tr{|\rho(t+\tau)-\rho(t)-\tau\mathcal{L}\rho(t)|},
  \label{eq:Dtime}
\end{equation}
where $\rho(t+\tau)$ is the density matrix containing the variational
parameters. Higher-order schemes exist as well, but constructing an
upper bound similar to Eq.~(\ref{eq:D}) requires to consider
higher-order correlations due to multiple applications of the
Liouvillian to the density matrix. A good compromise is the implicit
midpoint method, which is exact up to second order in $\tau$ while
only requiring a single application of the Liouvillian
\cite{Overbeck2016}.

\subsection{Comparison with mean-field methods}

For equilibrium problems, the variational method based on product
states is exactly equivalent to a mean-field decoupling of the
interaction terms. Remarkably, this is not the case for open quantum
systems. Within the mean-field approach to open systems
\cite{Tomadin2010,Diehl2010a}, a set of effective single site master
equations is considered that is obtained by tracing out the rest of the
system. For the $i$th site, the mean-field master equation reads
\begin{equation}
  \frac{d}{dt}\rho_i = \tr[\not{i}]{\frac{d}{dt}\rho} = -i[H_i^{MF},\rho_i] + \mathcal{D}_i(\rho_i),
  \label{eq:mf}
\end{equation}
where $H_i^{MF}$ and $\mathcal{D}_i$ are the mean-field Hamiltonian
and the mean-field dissipators, respectively. This set of equation is
then solved self-consistently, while for translationally invariant
systems it is often sufficient to consider an effective single site
problem.

\begin{figure}[t]
\includegraphics{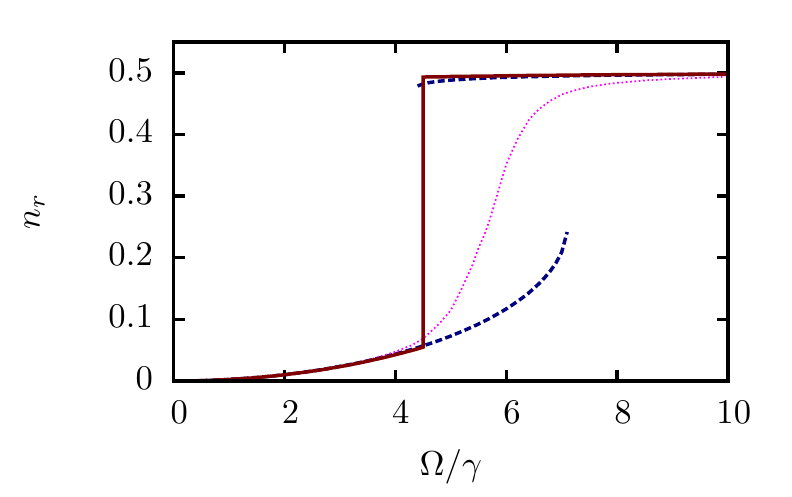}
\caption{Comparison of the solutions according to the variational
  principle (solid), the mean-field decoupling (dashed), and
  wave-function Monte-Carlo simulations for $4\times 4$ lattices for
  the up-spin density $n_r$ of the dissipative Ising model. The
  mean-field solution displays a region of bistability, while the
  variational solution correctly predicts a first order
  transition. From \cite{Weimer2015a}.}
\label{fig:mf}
\end{figure}

Due to the nonlinear structure of the mean-field equations of motion,
it is possible to have two or more independent solutions for the
steady state \cite{Lee2011}, see Fig.~\ref{fig:mf}. This also occurs
within mean-field theory for equilibrium systems close to first order
transitions. However, there one can always resort to the free energy,
which has to be minimal in thermal equilbrium. Unless one invokes the
variational principle, one cannot decide which of the solutions of
mean-field theory are stable and which ones are not. Interestingly,
the solution according to the variational principle and mean-field
theory are only identical in the limit of infinite dimensions, where
both approaches become exact \cite{Weimer2015}.

Mean-field theory predicts bistability for a wide range of models,
including the dissipative Ising model
\cite{Lee2011,Lee2012,Marcuzzi2014} or extended spin models
\cite{Parmee2018}, as well as driven-dissipative Bose-Hubbard models
\cite{LeBoite2013,Jin2013,LeBoite2014,Mertz2016}. So far, mean-field
bistability has been found in the absence of symmetries in the
underlying master equation, i.e., the two solutions are not connected
by a symmetry transformation. These properties have led to speculation
that bistability could be a genuine nonequilibrium phase, which has
stimulated several investigations whether this could indeed be the
case. However, the results of these investigation have all been
negative so far. Specifically, the variational principle predicts that
bistability is replaced by a first order transition both in the
dissipative Ising and in the driven-dissipative Bose-Hubbard model
\cite{Weimer2015}. For the dissipative Ising model, the existence of
the first order transition has been confirmed in tensor network
simulations, where bistability is found for low bond dimensions, but a
first order transition appears for higher bond dimensions
\cite{Kshetrimayum2017}, see Sec.~\ref{sec:tn2d}. In the case of the
driven-dissipative Bose-Hubbard model, the first order transition has
also been found in a field-theoretic treatment based on the Keldysh
formalism \cite{Maghrebi2016}, again confirming the variational
prediction. These results underscore that the conventional argument of
mean field theory becoming qualitiatively correct if the spatial
dimension becomes large enough appears to be incorrect for open
quantum systems. On the other hand, this argument seems to be much
more justified when applied to the variational principle (especially,
when considering the connection to equilibrium statistical physics
through the existence of the dissipative Ginzburg-Landau theory of
Eq.~(\ref{eq:gl})), however, even there one may have possible
counterexamples \cite{Mesterhazy2017}, which are not yet fully
understood.

\hw{Nevertheless, these findings do not rule out genuine bistability
  in open quantum systems per se, but only that such mean-field
  results need to be taken with caution. Classical models exhibiting
  extended coexistence regions \cite{Munoz2005} might still exhibit
  bistability after including quantum fluctuations. The situation is
  similar when it comes to limit cycles of open quantum many-body
  systems \cite{Chan2015}, which have been predicted to exist in
  sufficiently high-dimensional systems \cite{Owen2018}.}

One systematic extension of mean-field theory is cluster mean-field
theory, where the trace in Eq.~(\ref{eq:mf}) is not carried out over
all but one site but results in a larger cluster that has again to be
solved self-consistently \cite{Jin2016}. This strategy is in close
analogy to the cluster mean-field theory for statistical mechanics and
ground state problems \cite{Bethe1935,Oguchi1955}. Essentially,
cluster mean-field approaches treat the short-range physics more
accurately than bare mean-field theory, leading to better quantitative
estimates for phase transitions. However, the qualitative limitations
of bare mean-field theory remains, as these are resulting of
long-range fluctuations in the system. For open quantum many-body
models, cluster mean-field theory has been used to calculate the phase
diagram of the dissipative Heisenberg model given by
Eq.~\ref{eq:dheisen} \cite{Jin2016} and dissipative Ising models with
and without a $\mathbb{Z}_2$ symmetry \cite{Jin2018}.

Finally, it is also possible to systematically go beyond the
mean-field approximation using open system dynamical mean-field theory
(DMFT). DMFT is a mapping of a many-body lattice model onto a single
impurity problem that has to be solved in a self-consistent way
\cite{Georges1996}. \hw{Within DMFT, the approach is to start with an effective dynamical Green's function $\mathcal{G}_0$, which serves as a time-dependent version of a mean-field coupling. Considering the Fermi-Hubbard model as an example, $\mathcal{G}_0$ can used to express the effective action of a single site as
\begin{align}
  S_\text{eff} =& -\int\limits_0^\beta d\tau\int\limits_0^\beta d\tau'\sum\limits_\sigma f_\sigma^\dagger(\tau)\mathcal{G}_0^{-1} f_\sigma(\tau')\nonumber\\
  & + U\int\limits_0^\beta f^\dagger_\uparrow f^{\phantom{\dagger}}_\uparrow f^\dagger_\downarrow f^{\phantom{\dagger}}_\downarrow,
\end{align}
where $f_\sigma$ annihilates a fermion with spin $\sigma$, $\beta$ is
the inverse temperature, and $U$ is the on-site interaction. The
central idea of DMFT is to consider a self-consistent solution that
repoduces the dynamical Green's function $\mathcal{G}_0$. This constraint is satisfied by the solution to the DMFT equations for the local Green's function $G_0$, the dynamical Green's function $\mathcal{G}_0$, and the self-energy $\Sigma$ evaluated at the Matsubara frequencies $\omega_n = (2n+1)\pi/\beta$,
\begin{align}
  G_0(i\omega_n) &= \langle c_\sigma(i\omega_n)c_\sigma^*(i\omega_n)\rangle_{S_\text{eff}}\\
  G_0(i\omega_n) &= \left[ \mathcal{G}_0(i\omega_n)^{-1} - \Sigma(i\omega_n)\right]^{-1}\\
  G_0(i\omega_n) &= \int d\epsilon \frac{N(\epsilon)}{i\omega_n + \mu - \Sigma(i\omega_n) - \epsilon},
\end{align}
where $\mu$ is the chemical potential and $N(\epsilon)$ is the density of states \cite{Kollar2011}. The first step in bringing DMFT to open systems has been to use} effective Lindblad master equations
have to describe quantum transport in \emph{closed} quantum
systems using DMFT
\cite{Arrigoni2013,Titvinidze2015,Titvinidze2016}. Recently, this
approach has been extended to the case where already the initial
many-body problem describes an open quantum system \cite{Panas2018}.

\hw{A different method to systematically extend mean-field theory is
  to use projection operator methods. The central idea is to consider
  a single site of the many-body problem, with the rest of the system
  forming a non-Markovian environment. This non-Markovian master
  equation is then solved using standard projection operator
  techniques such as the Nakajima-Zwanzig method or the
  time-convolutionless master equation \cite{Breuer2002}. Initially,
  this approach has been used to describe the relaxation dynamics of
  local observables in a closed quantum system \cite{Weimer2008a},
  which has later been extended to the Lindblad dynamics of open
  systems \cite{Degenfeld-Schonburg2014}. There, the initial step is
  to introduce corrections $\Delta\mathcal{L}$ to the mean-field
  Liouvillian $\mathcal{L}_\text{MF}$ are introduced according to
  \begin{equation}
    \mathcal{L} = \mathcal{L}_\text{MF} + \Delta\mathcal{L}.
  \end{equation}
  The projection $\mathcal{P}$ removes all correlations and projects the system onto a product state, i.e.,
  \begin{equation}
    \mathcal{P}\rho = \prod\limits_i \rho_i.
  \end{equation}
  If the initial state at time $t_0$ is also a product state, the projected Lindblad master equation may be formally written as
  \begin{equation}
    \mathcal{P}\frac{d}{dt}\rho(t) = \mathcal{L}_\text{MF}\mathcal{P}\rho(t) + \mathcal{P} \Delta \mathcal{L} \int\limits_0^t dt' \mathcal{K}(t, t')\mathcal{P}\rho(t'),
  \end{equation}
  where the generator $\mathcal{K}$ has been introduced
  \cite{Degenfeld-Schonburg2014}. The generator $\mathcal{K}$ may then
  be expanded in terms of a power series of the beyond mean-field
  corrections $\Delta\mathcal{L}$. This projection operator approach has been used to investigate both dissipative XY models  \cite{Degenfeld-Schonburg2014} and the dissipative Heisenberg model \cite{Owen2018}. Remarkably, in the latter case a limit cycle behavior has been reported, which for sufficiently large spatial dimensions also survives under inclusion of the terms beyond mean-field. Consequently, it would be interesting to learn whether the projection operator approach is also capable to correctly identify the replacement of mean-field bistability by a first order transition in the dissipative Ising model.}

	\subsection{Variational tensor network methods}

        \label{sec:vartensor}

        Given the successes of tensor network methods discussed in
        Sec.~\ref{sec:tensor}, it appears natural to combine them with
        variational methods for the study of open quantum many-body
        systems. However, the main challenge is that the natural trace
        norm for constructing the variational principle cannot be
        calculated efficiently in a tensor network
        respresentation. This has led to the use of different norms as
        possible alternatives \cite{Cui2015,Mascarenhas2015}, see
        Sec.~\ref{sec:tensor1d}.

        On the one hand, the choice of the norm is not really relevant
        if the value of the norm is very low (i.e., comparable to the
        machine precision of the numerical simulation), as then the
        solution is almost exact from any point of view. On the other
        hand, choosing a non-natural norm is a potential source of errors
        that is not under control of the variational algorithm. In
        practice, this difficulty will mostly manifest itself for
        higher-dimensional problems, as there the bond dimensions that
        can be reached are severly constrained by the compuational
        resources \cite{Kshetrimayum2017}. But even for
        one-dimensional systems, there are computationally challenging
        problems involving long relaxation times \cite{Carollo2019},
        where an arbitrarily low variational norm might not be
        reachable.

        A way out of this problem can be realized by representing the
        density matrix in terms of an ensemble of pure states and use
        a variational tensor network formulation for these pure states
        \cite{Transchel2014}. In this case, the density matrix is
        parametrized according to
        \begin{equation}
          \rho = \int p(\alpha,\bar{\alpha})\ketbra{\psi(\alpha)}d\alpha d\bar{\alpha},
        \end{equation}
        where $\ket{\psi(\alpha)}$ is a variational wave function with
        variational parameters $\alpha$ and $p(\alpha,\bar{\alpha})$
        is the associated probability distribution. Crucially, the
        variational norm associated with the effective Hamiltonian of
        the master equation $H_\text{eff} = H - i/2 \sum_i c_i^\dagger
        c_i$ can now be calculated as
        \begin{equation}
          D_H = |H_\text{eff}\ket{\psi(\alpha)}|^2.
        \end{equation}
        This expression can both be computed efficiently using tensor
        network methods and corresponds to the natural trace norm when
        evaluated over the full ensemble. The quantum jump terms of
        the master equation can be treated in a similar fashion
        \cite{Transchel2014}.
	

        \subsection{Variational quantum Monte-Carlo methods}

        The central idea behind quantum Monte-Carlo methods is to
        rewrite a quantum many-body problem in terms of a sampling
        over a classical probability distribution
        \cite{Batrouni2011}. However, the existence of destructive
        interference in quantum mechanics can lead to corresponding
        classical probabilities that are negative, which is the root
        of the famous sign problem. One common workaround is to sample
        over the absolute value of the probability distribution
        instead, but this comes at the price of the complexity of the
        computation increasing exponentially with the system size
        \cite{Troyer2005}. Open quantum many-body systems are
        especially prone to the sign problem since the eigenvalues of
        the Liouvillian can even be complex
        \cite{Nagy2018}. Nevertheless, Monte-Carlo sampling can be
        useful even in the presence of the sign problem, if the
        required resources for the Monte-Carlo sampling are lower than
        for a full solution of the problem.

        The first quantum Monte-Carlo simulation of an open quantum
        many-body problem has been based on a non-variational
        full-configuration-interaction Monte Carlo algorithm
        \cite{Nagy2018}, which is better equipped to deal with the
        sign problem without completely negating it. For the
        magnetization of a dissipative XYZ model on small lattices,
        the quantum Monte-Carlo simulation is in excellent agreement
        with wave-function Monte-Carlo results.

\begin{figure}[t]
          \includegraphics[width=0.8\linewidth]{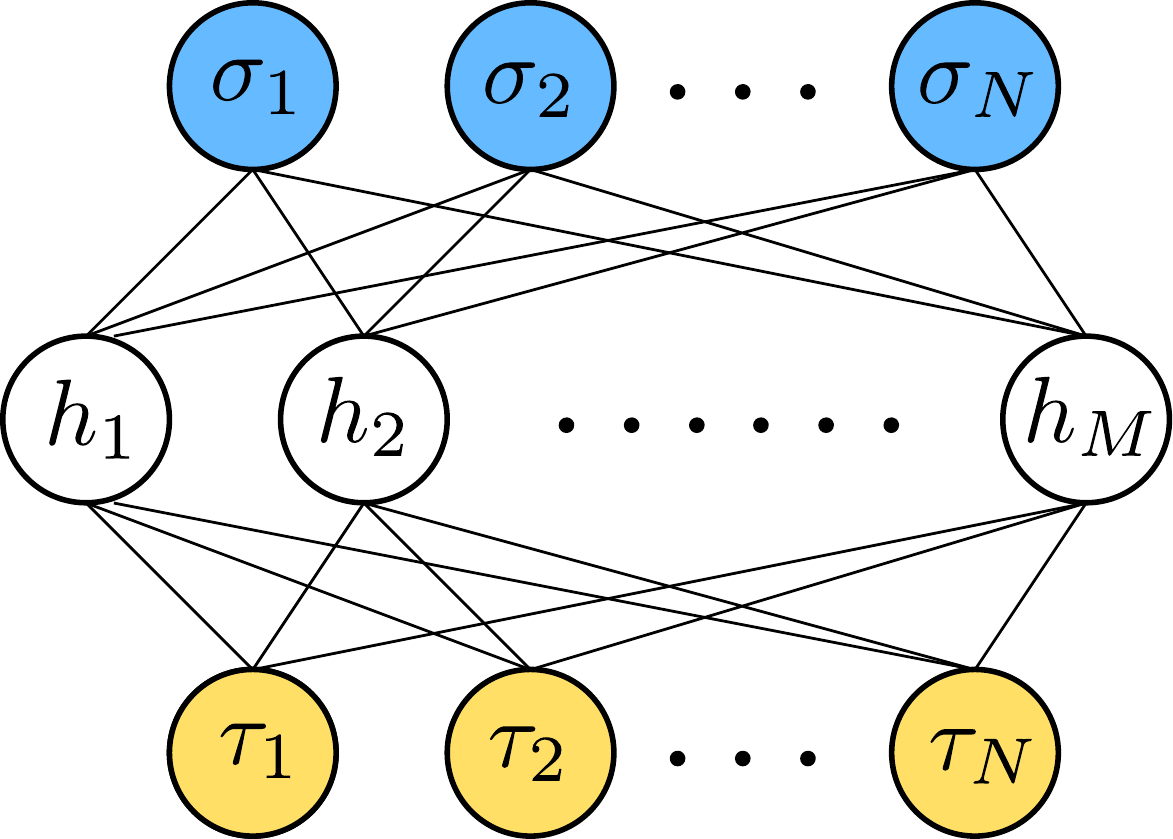}
          \caption{Node structure of a restricted Boltzmann machine
            for open quantum systems. The vectorized density matrix is
            realized in terms of a physical layer $\sigma_i$,
            corresponding to a set of spin $1/2$ variables. These are
            coupled to the nodes of a hidden layer $h_i$, which are
            again coupled to the third layer $\tau_i$, which
            represents the adjoint of the physical layer.}
          \label{fig:rbm}

        \end{figure}

        Recently, variational Monte-Carlo methods have been applied to
        open quantum systems
        \cite{Hartmann2019,Yoshioka2019,Nagy2019,Vincentini2019}. These
        approaches are inspired by using variational wave function
        corresponding to restricted Boltzmann machines (RBMs)
        \cite{Carleo2017}, which were first introduced in the context
        of neural network simulations. The main idea behind RBM wave
        functions is shown in Fig.~\ref{fig:rbm}, where an additional
        hidden layer introduces variational parameters associated with
        the quantum correlations of the many-body
        state. The entries of the vectorized density matrix are then given by
        \begin{align}
          _\sharp\braket{\mathbf{\sigma},\mathbf{\tau}}{\rho}_\sharp &= \frac{1}{Z}
            \sum\limits_{\{h_j\}} \exp\left(\sum\limits_{ij} W_{ij} \sigma_ih_j + W_{ij}^*\tau_ih_j\right)\nonumber\\
            &\times \exp\left(\sum\limits_i a_i \sigma_i + a_i^* \tau_i + \sum\limits_j b_j h_j\right),
        \end{align}
        where the $W_{ij}$, $a_i$, and $h_j$ are variational
        parameters. Interestingly, there is a close connection between
        RBM wave functions and matrix product states
        \cite{Deng2017,Chen2018a}, however, RBMs are potentially also
        capable to describe long-range entangled quantum states.

        For the variational Monte-Carlo samplings, different norms
        have been put forward. One possibility is to consider the
        Hilbert-Schmidt norm of the time evolution \cite{Hartmann2019}
        or the steady state \cite{Vincentini2019}. Interestingly, in
        the latter case, the variational norm $D$ has been normalized
        according to the purity $\trtxt{\rho^2}$, i.e.,
        \begin{equation}
          D = \frac{\tr{\dot\rho^2}}{\tr{\rho^2}}.
        \end{equation}
        This norm is not biased towards the maximally mixed state as
        mentioned above. An alternative approach to construct a
        suitable norm is to mimize the Hermitian
        $\mathcal{L}^\dagger\mathcal{L}$ in close analogy to a ground
        state problem \cite{Yoshioka2019}. Finally, it is possible to
        consider the equivalent of an expectation value for vectorized
        density matrices according to $_\sharp\langle
        \rho|\mathcal{L}_\sharp|\rho\rangle_\sharp/_\sharp\langle
        \rho|\rho\rangle_\sharp$ \cite{Nagy2019}. With the respect to
        the more natural trace norm for density matrices, the RBM
        approaches behave similarly to the tensor network simulations
        discussed in Sec.~\ref{sec:vartensor}. However, since RBMs can
        be applied to two-dimensional models in a straightforward way,
        it will be very interesting to see how these methods perform
        for the investigation of dissipative phase transitions, in
        particular in critical systems.

	\section{Phase space and related methods}
	Other methods have also been used with relative success in the study of open quantum systems, such as phase space methods, as well as methods based on hierarchy equations. In this section we explain two of such examples, namely, truncated Wigner approximations and Bogoliubov-Born-Green-Kirkwood-Yvon (BBGKY) hierarchies. The resulting methods are very general in purpose, and can be applied to a wide variety of systems, yet in what follow we discuss concrete examples. 
	
	\subsection{Truncated Wigner approximation}
	In the context of phase-space related methods, truncated Wigner approximations \ro{were first used in \cite{Carusotto2005} to driven-dissipative microcavity polariton system coherently driven into the optical parametric oscillator regime, also reviewed in \cite{Carusotto2013} and revisited in \cite{Dagvadorj2015} as an example of a 2d driven-dissipative non-equilibrium phase transition.} The Hamiltonian for the system is given by 
	\beq
	H_S = \int d\vec{r} \left( \psi_X^\dagger ~ \psi_C^\dagger \right) 
	\begin{pmatrix}
		\frac{ - \nabla^2}{2m_X} + \frac{g_X}{2} |\psi_X|^2 & \frac{\Omega_R}{2} \\
		\frac{\Omega_R}{2} & \frac{- \nabla^2}{2 m_C} 
	\end{pmatrix}
	\begin{pmatrix} 
		\psi_X \\
		\psi_C
	\end{pmatrix},
	\eeq
	with cavity and photon field operators $\psi_{X,C}(\vec{r}, t)$, spatial coordinate $\vec{r} = (x,y)$, $m_{X,C}$ the exciton and photon masses, $g_X$ the exciton-exciton interaction strength, and $\Omega_R$ the Rabi splitting. One introduces the effect of an external drive (pump) as well as incoherent decay by adding a system-bath Hamiltonian given by 
	\beqa
	H_{SB} &=& \int d \vec{r} \left( F(\vec{r},t) \psi^\dagger_C(\vec{r},t) + h.c. \right) \nonumber \\
	&+& \sum_{\vec{k}} \sum_{l = X, C} \left( \xi_{\vec{k}}^l \left( \psi_{l,\vec{k}}^\dagger(t) B_{l,\vec{k}} + h.c. \right) + \omega_{l,\vec{k}} B^\dagger_{l,\vec{k}} B_{l,\vec{k}}\right), \nonumber \\
	\eeqa
	with $\psi_{l,\vec{k}}(t)$ the Fourier transform of the field operators in real space, $ B_{l,\vec{k}}$ and $ B_{l,\vec{k}}^\dagger$ the bath's bosonic anihilation and creation operators with energy $\omega_{l,\vec{k}}$, which describes the decay for both excitons and cavity photons. The decay is compensated by an external homogeneous coherent pump $F(\vec{r},t) = f_p e^{i(\vec{k}_p \cdot \vec{r} - \omega_p t)}$, injecting polaritons with momentum $\vec{k}_p$ and energy $\omega_p$. 
	
	By using standard quantum optical methods, one can trace out the bath within the Markovian approximation and obtain a Master equation for the system. There is, however, an alternative approach by means of phase-space techniques. In particular, one can represent the quantum fields as quasiprobability distribution functions. The Fokker-Planck partial differential equation that governs the dynamics of such distributions can be mapped to a stochastic differential equation, which can be solved using different techniques. For the example that we are discussing, one solves the equation on a finite grid with lattice spacing $a$. The most suitable quasiprobability distribution for this example is the Wigner representation, which is also the most suitable one for numerical implementation. By truncating the corresponding Fokker-Planck equation in the limit $(g_X/\kappa_{X,C} a^2) \ll 1$, with $\kappa_{X,C}$ the exciton and photon decay rates, and keeping up to second-order derivatives only, one obtains the stochastic differential equation 
	\beq
	i d 
	\begin{pmatrix}
		\psi_X  \\
		\psi_C 
	\end{pmatrix}
	= \left( H^\prime_{HF} 
	\begin{pmatrix}
		\psi_X \\
		\psi_C 
	\end{pmatrix}
	+
	\begin{pmatrix}
		0 \\
		F
	\end{pmatrix}
	\right) dt + i
	\begin{pmatrix}
		\sqrt{\kappa_X} dW_X \\ 
		\sqrt{\kappa_C} dW_C 
	\end{pmatrix}. 
	\eeq
	In this equation, $dW_{l=X,C}$ are Wiener noise terms, and $H^\prime_{MF}$ is given by 
	\beq
	H^\prime_{MF} = 
	\begin{pmatrix}
		\frac{ - \nabla^2}{2m_X} + {g_X} \left(|\psi_X|^2 - \frac{1}{a^2} \right) - i \kappa_X & \frac{\Omega_R}{2} \\
		\frac{\Omega_R}{2} & \frac{- \nabla^2}{2 m_C} - i \kappa_C
	\end{pmatrix}
	\eeq
	The resulting stochastic differential equation can then be solved using standard methods and softeare packages for this purpose. 	
	
\ro{Let us add that recently, this method has also been used to study critical slowing down in photonic lattices \cite{Vicentini2018}, as well as extended to disordered quantum many-body system (the so-called optical stochastic unraveling for disordered systems) \cite{Vicentini2019}. }
	
	\subsection{BBGKY hierarchy equations}
	
	It is also possible to study open quantum systems via the so-called Bogoliubov-Born-Kirkwood-Yvon hierarchy \cite{Liboff2003}. In a nutshell, this is a hierarchy of equations aimed to describe a system of a large number of interacting particles. As such, the idea is very generic. But as shown in \cite{Navez2010}, it can also be applied directly in the context of open dissipative systems in order to obtain a hierarchy of equations for the different reduced density matrices. 
\clearpage	
	The way this approach works is quite intuitive. Consider the reduced density matrices for one lattice site $\rho_\mu$, for two lattice sites $\rho_{\mu \nu}$, and so on. We separate the correlated parts as $\rho_{\mu \nu} = \rho_{\mu \nu}^c + \rho_\mu \rho_\nu$, as well as $\rho_{\mu \nu \lambda} = \rho_{\mu \nu}^c \rho_\lambda + \rho_{\mu \lambda}^c \rho_\nu + \rho_{\nu \lambda}^c \rho_\mu + \rho_\mu \rho_\nu \rho_\lambda$, and so on. The method that will be discussed in what follows in based on the scaling hierarchy of correlations 
	\beq
	\rho_{\mathcal{S}}^c = O\left( Z^{1 - |\mathcal{S}|} \right), 
	\label{hierar}
	\eeq
	with $|\mathcal{S}|$ the number of lattice sites in set $\mathcal{S}$. The different reduced density matrices can also be computed using the generating functional $\mathcal{F}(\alpha_\mu) = \log {\rm Tr}(\rho \prod_\mu (\mathbb{I}_\mu + \alpha_\mu))$, with $\alpha_\mu$ an arbitrary operator acting on site $\mu$. Using such a functional one has $\rho_\mu = \partial \mathcal{F} / \partial \alpha_\mu |_{\alpha = 0}$, as well as $\rho_{\mu \nu}^c = \partial^2 \mathcal{F} / \partial \alpha_\mu \partial \alpha_\nu |_{\alpha = 0}$, and so on.  Next, the Liouville operators $\mathcal{L}_\mu$ and $\mathcal{L}_{\mu \nu}$ acting on one and two sites are introduce via the dissipation equation $i \partial_t \rho = [H, \rho] + \sum_\mu \mathcal{L}_\mu \rho + \sum_{\mu \nu} \mathcal{L}_{\mu \nu} \rho / Z$, with $Z$ the coordination number of the Hamiltonian (e.g., the number of tunneling neighbours at any given site for a Hubbard-like Hamiltonian). Following these equations, the time evolution of $\mathcal{F}$ is given by 
	\begin{widetext}
		\beq
		i \frac{\partial}{\partial t} \mathcal{F}(\alpha) = \sum_\mu {\rm Tr}_\mu \left( \alpha_\mu \mathcal{L}_\mu \frac{\partial \mathcal{F}}{\partial \alpha_\mu} \right) + \frac{1}{Z} \sum_{\mu \nu} {\rm Tr}_{\mu \nu} \left( (\alpha_\mu + \alpha_\nu + \alpha_\mu \alpha_\nu ) \mathcal{L}_{\mu \nu} \left( \frac{\partial^2 \mathcal{F}}{\partial \alpha_\mu \partial \alpha_\nu} + \frac{\partial \mathcal{F}}{\partial \alpha_\mu} \frac{\partial \mathcal{F}}{\partial \alpha_\nu} \right) \right). 
		\eeq 
	\end{widetext}
	Using this equation, one can take derivatives and obtain a set of equations for the correlated density matrices, 
	\begin{widetext}
		\beqa
		i \frac{\partial}{\partial t} \rho_{\mathcal{S}}^c  &=& \sum_{\mu \in \mathcal{S}} \mathcal{L}_\mu \rho_{\mathcal{S}}^c + \frac{1}{Z} \sum_{\mu \nu \in \mathcal{S}} \mathcal{L}_{\mu \nu} \rho_{\mathcal{S}}^c + \frac{1}{Z} \sum_{k \notin \mathcal{S}} \sum_{\mu \in \mathcal{S}} {\rm Tr}_k \left( \mathcal{L}_{\mu k}^{{S}} \rho_{\mathcal{S} \cup k}^c + \sum_{\mathcal{P}  \subseteq \mathcal{S} \setminus \{ \mu \}}^{\mathcal{P} \cup \bar{\mathcal{P}} = \mathcal{S} \setminus \{ \mu \}} \mathcal{L}_{\mu k}^{{S}} \rho_{\{ \mu \} \cup \mathcal{P}}^c \rho_{\{ k \} \cup \mathcal{\bar{P}}}^c \right) \nonumber \\
		&+& \frac{1}{Z} \sum_{\mu \nu \in \mathcal{S}} \sum_{\mathcal{P} \subseteq \mathcal{S} \setminus \{ \mu, \nu \}}^{\mathcal{P} \cup \mathcal{\bar{P}} = \mathcal{S} \setminus \{ \mu, \nu \}} \left( \mathcal{L}_{\mu \nu} \rho_{\{ \mu \} \cup \mathcal{P}}^c \rho_{\{ \nu \} \cup \mathcal{\bar{P}}}^c - {\rm Tr}_\nu \left( \mathcal{L}_{\mu \nu}^{{S}} \left( \rho_{\{ \mu, \nu \} \cup \mathcal{\bar{P}}}^c + \sum_{\mathcal{Q} \subseteq \mathcal{\bar{P}}}^{\mathcal{Q} \cup \mathcal{\bar{Q}} = \mathcal{\bar{P}}}  \rho_{\{ \mu \} \cup \mathcal{Q}}^c \rho_{\{ \nu \} \cup \mathcal{\bar{Q}}}^c \right) \right) \rho_{\{ \nu \} \cup \mathcal{P}}^c \right) 
		\eeqa
	\end{widetext}
	with $\mathcal{L}_{\mu \nu}^S = \mathcal{L}_{\mu \nu} + \mathcal{L}_{\nu \mu}$. This hierarchy of equations for the reduced density matrices is preserved in time. Moreover, it allows us to write explicit equations for the one- and two-site density matrices. For the one-site matrix one gets
	\beq
	i \frac{\partial}{\partial t} \rho_\mu = \mathcal{L}_\mu + \frac{1}{Z} \sum_k {\rm Tr}_k \left( \mathcal{L}_{\mu k}^S \left( \rho_{\mu k}^c + \rho_\mu \rho_k \right) \right), 
	\eeq
	and for the two-site matrix one has
	\beqa
	i \frac{\partial}{\partial t} \rho_{\mu \nu} &=& \mathcal{L}_\mu \rho_{\mu \nu}^c \frac{1}{Z} \mathcal{L}_{\mu \nu} \left( \rho_{\mu \nu}^c + \rho_\mu \rho_\nu \right) \nonumber \\
	&+& \frac{1}{Z} \sum_{k \neq \mu, \nu} {\rm Tr}_k \left( \mathcal{L}_{\mu k}^S \left( \rho_{\mu \nu k}^c + \rho_{\mu \nu}^c \rho_k + \rho_{\nu k}^c \rho_\mu \right) \right) \nonumber \\
	&-& \frac{\rho_\mu}{Z} {\rm Tr}_\mu \left( \mathcal{L}_{\mu \nu}^S \left( \rho_{\mu \nu}^c + \rho_\mu \rho_\nu \right) \right) + (\mu \leftrightarrow \nu). 
	\eeqa
	By combining the above expressions with Eq.(\ref{hierar}), one can expand in powers of $1/Z$ and obtain different approximations for the one- and two-particle behaviour. 
	
	This approach can be implemented for a variety of systems (spins, bosons, fermions...) and has the advantage of being independent of the dimensionality of the system. For instance, in \cite{Navez2010} it was applied to a lattice Bose-Hubbard model. The method can be used to obtain analytical expansions, as well as to facilitate efficient numerical simulations.
	
	\ro{\section{Linked cluster expansion methods}}
	
	\ro{Methods based on linked-cluster expansions have also been recently put forward in the study of open quantum many-body systems, so far focusing on the study of two-dimensional spin systems with incoherent spin relaxation \cite{Biella2018}. The method numerically targets expectation values of observables in the steady state (at long times) of the master equation.}
	
	\ro{Mathematically, the procedure is as follows: let us assume (without loss of generality) that the Liouvillian can be expanded as a sum of two-body terms, i.e.,
		\beq
		\mathcal{L} = \sum_{\langle i, j \rangle}\alpha_{ij} \mathcal{L}_{ij},
		\eeq
		with $\alpha_{ij}$ some local coupling strength. For the sake of simplicity let us define $k \equiv (i,j)$ as a combined index. The expectation value $O$ of an observable $\hat{O}$ can be expanded in terms of powers of $\alpha_k$, i.e., 
		\beq
		O(\{ \alpha_k \}) = \sum_{\{ n_k \}} O_{\{ n_k \}} \prod_k \alpha_k^{n_k}, 
		\eeq
		with $n_k$ running over all non-negative integers for all $k$. It is clear that all possible polynomials in $\alpha_k$ are included in the above expression, which can be reorganized in clusters as follows: 
		\beq
		O = \sum_c W_{[O]}(c), 
		\eeq
		with $c$ a non-empty set of $k$-indexes identifying the sites belonging to the cluster. The cluster weight $W_{[O]}(c)$ contains all the terms in the expansion with at least one power of $\alpha_k$, for all $k$ in $c$, and no powers of $\alpha_k$ of $k$ does not belong to $c$. These terms obey the recurrence relation 
		\beq
		W_{[O]}(c) = O(c) - \sum_{s \subset c} W_{[O]}(s), 
		\eeq
		with 
		\beq
		O(c) = {\rm Tr} (\hat{O} \rho_s(c))
		\eeq
		being the expectation value of the observable in the steady-state $\rho_s(c)$ for the finite cluster $c$. Taking into account symmetries in the system, the expectation value per site in the thermodynamic limit can be written as 
		\beq
		\frac{O}{L} = \sum_{n = 1}^{\infty} \left( \sum_{c_n} l(c_n) W_{[O]}(c_n) \right),
		\eeq
		with $L \rightarrow \infty$ the size of the system, the outer sum running over all possible cluster sizes $n$, and the inner sum over all topologically different clusters $c_n$ of size $n$, with $l(c_n)$ their multiplicity. This series expansion can be truncated up to a cluster size $R$, thus giving rise to a plausible approximation method also valid for open systems.}

        \hw{The linked cluster expansion works very well for the dissipative Heisenberg model \cite{Biella2018}, where an exact product state solution can be used as a starting point of the expansion. In this case, it is even possible to calculate phase boundaries and critical exponents of a dissipative phase transition between a paramagnet and a ferromagnet. The situation is quite different for the dissipative Ising model, where the expansion series failed to converge even for a 10$^\text{th}$ order expansion \cite{Jin2018}.}
	
	\bigskip 

	\section{Summary and outlook}

        The enormous effort to develop novel simulation methods to
        investigate open quantum many-body systems has enabled us to
        review a large variety of numerical methods. \ro{To be more specific, in this review we considered methods for the Markovian quantum master equation (assuming a weak-coupling limit), including mean-field, stochastic  methods, tensor networks, variational methods, quantum Monte-Carlo, truncated Wigner approximation, BBGKY hierarchy equations, and linked cluster expansions.} While so far, no
        method has emerged that is universally optimal for all cases,
        there have been several very promising
        developments \ro{with different methods for different regimes.}  \ak{Even with such major technical advances discussed in this review, there are still many open problems which are inacessible with these state-of-the-art numerical techniques. To give concrete examples of actual physical problems, one may consider a very common setting in the context of Rydberg atoms where the interaction is often long-ranged and cannot be approximated with just a nearest-neighbour Hamiltonian \cite{Schachenmayer_2015,Labuhn2016,Browaeys_2016}. Even TN techniques will face a difficult challenge specially in 2D while encountering such problems although there has been promising developments even in this direction recently \cite{2019arXiv191104592O}. Other challenging problems include the existence of AF order in 3D dissipative Ising model which is an open question that appears hard to answer. This is again relevant to ongoing experiments with Rydberg atoms, which one cannot reliably simulate at the moment \cite{Carr2013,Malossi2014,Helmrich2018}. Phase transitions and universality classes of dissipative models is another class of problem which has proven to be quite difficult for numerical techniques \cite{Diehl2010,Carmichael2015,Fink2017,Biondi2017}.}

        Certainly, the largest confidence in a
        simulation result can be achieved if it is reproducible using
        a complementary simulation approach. Despite these caveats,
        one can make several key observations about the particular
        methods covered in this review. The first observation is that
        mean-field methods are considerably less reliable for open
        system than their counterparts for closed systems, although
        the reason for this discrepancy is still an open
        question. Furthermore, tensor network methods have
        demonstrated their ability to successfully tackle many hard
        problems surrounding open many-body systems and resolve
        long-standing open questions. \ro{A particularly interesting and promising case is that of open 2d systems, which is unexplored territory to a great extent.} As for the variational methods
        discussed in this review, there appears to be a tradeoff
        between the formal suitability of the norm and its efficient
        computability. It will be interesting to see if and how this
tradeoff will be resolved in future work. \ak{We provide a summary in Tab.~\ref{tab2} comparing the different techniques we have discussed above.}
        \begin{table*}
        	\centering
        	\begin{tabular}{||c|c|c|c|c|c|c||} 
        		\hline 
        		~~ & ~~ WFMC~ & ~TN~ &~Variational Principle~&VQMC & CMF & TWA~  \\
        		\hline 		 System size &20 & TDL & TDL & 16 & TDL & 400 \\
        		 Dimensions & 1D,2D & 1D,2D & any\footnote{Works better in higher dimensions\label{fn1}} & any & any\footnote{Works better in higher dimensions} & any \\
        		Local Hilbert space & small & small & large & large & small & large \\ 
        		Fermionic systems & Yes & Yes & partially & No & partially & unknown \\
        		Inhomogeneous systems & good & good & bad & good & good & good \\
        		Critical exponents & good & good & good\footnote{For states with thermal statistics} & unknown & bad & unknown \\
        		\hline
        	\end{tabular}
        	\caption{\ak{Table comparing the different simulation methods discussed} \hw{in this review. We differentiate the methods by the system sizes that can be simulated, the spatial dimensions, contraints on the local Hilbert space dimension, whether fermionic systems can be treated, the simulation performance for inhomogeneous systems, and whether the correct critical exponents of phase transitions can be obtained.}}
        	\label{tab2}
        \end{table*}

        The progress in recent years in simulating open quantum
        systems has brought the field to a level where one has a wide
        range of tools at hand to systematically compare to
        experimental results, in particular in the context of quantum
        simulations. Combined with the experimental ease of preparing
        the steady state of an open quantum system, these are good
        reasons to believe that \ro{the study of strongly-correlated open quantum many-body systems will become a research topic with impact in other areas of science, such as material design and quantum computation.}

        \begin{acknowledgments}

          We thank C.~Ciuti for valuable feedback on our
          manuscript. This work was funded by the Volkswagen
          Foundation, by the Deutsche Forschungsgemeinschaft (DFG,
          German Research Foundation) within SFB 1227 (DQ-mat, project
          A04), SPP 1929 (GiRyd), and under Germany’s Excellence
          Strategy -- EXC-2123 QuantumFrontiers -- 390837967.

        \end{acknowledgments}

        \bibliography{open}

	
\end{document}